\documentclass[]{spie} 

\usepackage{amsmath,amsfonts,amssymb}
\usepackage{graphicx}
\usepackage[colorlinks=true, allcolors=blue]{hyperref}
\usepackage{lineno}
\usepackage{subcaption}
\usepackage{comment}

\providecommand{\sil}[1]{{#1}}

\title{Status and future development of the COSMOCal Project for absolute CMB polarization calibration}

\author[a]{S. Micheli$^1$}
\author[a]{A. Ritacco}
\author[b]{A. Carones}
\author[c]{J. Aumont}
\author[d]{S. Berta}
\author[e]{L. Bizzarri}
\author[f]{F. Boulanger}
\author[a]{A. Catalano}
\author[g]{F. Cuttaia}
\author[f]{A. Denis}
\author[h]{F.-X. D\'{e}sert}
\author[d]{A.-L. Fontana}
\author[i]{D. Gonz\'{a}lez Ovejero}
\author[a]{B. Kalemi}
\author[b]{N. Krachmalnicoff}
\author[d]{S. Leclercq}
\author[j]{T. Louis}
\author[a]{J.-F. Mac\'{i}as-P\'{e}rez}
\author[k]{B. Maffei}
\author[l]{T. Mcnamara}
\author[m]{M. Migliaccio}
\author[c]{L. Montier}
\author[f]{P. Morfin}
\author[l]{M. Moulin}
\author[k]{L. Mousset}
\author[n]{M. Murgia}
\author[o]{I. Myserlis}
\author[e]{F. Nati}
\author[n]{P. Ortu}
\author[f]{M. P\'{e}rault}
\author[p]{G. Pisano}
\author[n]{T. Pisanu}
\author[h]{N. Ponthieu}
\author[q]{S. Savorgnano}
\author[g]{L. Terenzi}
\author[r]{J. Treuttel}
\author[c]{L. Vacher}
\author[a]{C. Vescovi}
\author[e]{M. Zannoni}

\affil[a]{Universit\'{e} Grenoble Alpes, CNRS, LPSC Grenoble, France}
\affil[b]{International School for Advanced Studies (SISSA), Via Bonomea 265, Trieste 34136, Italy and INFN Sezione di Trieste, Via Valerio 2, Trieste 34127, Italy}
\affil[c]{Institut de Recherche en Astrophysique et Plan\'{e}tologie, CNRS-INSU, France}
\affil[d]{Institut de Radio-Astronomie Millim\'{e}trique, France}
\affil[e]{Department of Physics, University of Milano-Bicocca, Italy}
\affil[f]{Laboratoire de Physique de l'Ecole normale sup\'{e}rieure, ENS, Universit\'{e} PSL, CNRS, Sorbonne Universit\'{e}, Universit\'{e} Paris Cit\'{e}, F-75005 Paris, France}
\affil[g]{INAF - Osservatorio di astrofisica e scienza dello spazio Bologna, Italy}
\affil[h]{Univ. Grenoble Alpes, CNRS, IPAG, 38000 Grenoble, France}
\affil[i]{Institut d'Electronique et des Technologies du num\'{e}rique, France}
\affil[j]{Laboratoire de Physique des 2 infinis Ir\`{e}ne Joliot-Curie, CNRS, France}
\affil[k]{Institut d'Astrophysique Spatiale, CNRS-Universit\'{e} Paris-Saclay, France}
\affil[l]{Centre Spatial Universitaire de Grenoble, Universit\'{e} Grenoble Alpes, France}
\affil[m]{Dipartimento di Fisica, Universit\`{a} di Roma “Tor Vergata”, via della Ricerca Scientifica 1, I-00133, Roma, Italy}
\affil[n]{INAF - Osservatorio Astronomico di Cagliari, Italy}
\affil[o]{Institut di Radioastronomie Millim\'{e}trique, Avenida Divina Pastora, 7, Local 20, E--18012 Granada, Spain}
\affil[p]{Sapienza Universit\`{a} di Roma, Italy}
\affil[q]{Department of Physics, Boston University, United States}
\affil[r]{LERMA, Observatoire di Paris-PSL, CNRS, France}

\authorinfo{$^1$E-mail: silvia.micheli@lpsc.in2p3.fr}

\begin{document} 
\maketitle


\begin{abstract}
Cosmic Microwave Background (CMB) polarization measurements are pushing instrumental sensitivities to levels where calibration systematics become a dominant limitation. The large dynamic range between cosmological and Galactic emission prevents future experiments from relying only on diffuse sky measurements or standard celestial calibrators. To address this challenge, the COSmological Microwave Observations Calibrator (COSMOCal) project proposes an artificial calibration source deployed as a guest payload on a geostationary satellite, scheduled for launch by the Eutelsat group by 2030. This source will provide stable, well-characterized polarized microwave signals accessible to multiple ground-based observatories. In this work, we present the status of the project, the updated development timeline, and the refined scientific and technical requirements, defined with the observatories that plan to use this calibration source. Furthermore, we investigate the interplay between instrumental systematics and component separation in the presence of complex models of interstellar dust emission. We discuss in this paper how this can impact the recovery of the primordial signal, and whether residual calibration errors can degrade the performance of foreground cleaning algorithms.
\end{abstract}


\section{INTRODUCTION}
\label{sec:intro}  
The Cosmic Microwave Background (CMB) offers a unique opportunity to probe the physics of the early universe. Precise measurements of the CMB have firmly established the $\Lambda$CDM model as the standard cosmological model and strongly support the inflationary paradigm \cite{PlanckCollaboration:2020I-cosmology, PlanckCollaboration:2020X-inflation}, which predicts a phase of accelerated expansion right after the primordial singularity \cite{Guth:1982}. This rapid expansion left its signature on the CMB polarization field, which can be decomposed into two geometrically distinct components: the even-parity, gradient-like $E$-modes and the odd-parity, curl-like $B$-modes \cite{Kamionkowski:1997GW, Zaldarriaga:1997}. While $E$-modes have already been measured with impressive accuracy \cite{PlanckCollaboration:2020I-cosmology}, primordial $B$-modes are much more elusive. A key prediction of inflation is the existence of a background of primordial gravitational waves which would leave a unique imprint on the polarization field in the form of primordial $B$-modes \cite{Kamionkowski:2016}. Their amplitude is parametrized by the tensor-to-scalar ratio $r$, which directly connects to the energy scale at which inflation took place \cite{Lyth:1997}. To date, no primordial $B$-modes detection has been performed, and only upper limits have been set, the tightest being $r<0.032$ ($95\%$ CL) \cite{Tristram:2022}. Detecting this signal is thus one of the major goals of upcoming CMB experiments, such as \textit{LiteBIRD} \cite{Ghigna:2024} and the Simons Observatory (SO) \cite{Ade:2019}. However, reaching these unprecedented sensitivity levels makes experiments highly susceptible to instrumental systematic effects. In particular, a precise absolute calibration of the detectors polarization angles and beam profiles is required to minimize spurious $E$-to-$B$ leakage \cite{Hu:2003}. This leakage can not only mask the primordial $B$-mode signal, but it also degenerates with cosmic birefringence, a parity-violating effect that rotates the polarization plane of CMB photons, generating a non-zero $EB$ cross-correlation \cite{Minami:2020}. If the polarization angle of a detector has a mis-calibration offset $\Delta \alpha$, the observed $E$- and $B$-mode spherical harmonic coefficients can be expressed as a rotation of the intrinsic fields \cite{Kamionkowski:1997GW, Zaldarriaga:1997}:
\begin{equation}
    E^{\text{obs}}_{\ell m} = E_{\ell m} \cos(2\Delta \alpha) - B_{\ell m} \sin(2\Delta \alpha) ,
\end{equation}
\begin{equation}
    B^{\text{obs}}_{\ell m} = E_{\ell m} \sin(2\Delta \alpha) + B_{\ell m} \cos(2\Delta \alpha) .
\end{equation}
Given that the intrinsic $E$-mode amplitude is orders of magnitude larger than the expected $B$-mode signal, even a minor angular offset $\Delta \alpha$ induces significant leakage into the observed $BB$ angular power spectrum:
\begin{equation}
    C^{BB,\text{obs}}_{\ell} =  C^{EE}_{\ell} \sin^2(2\Delta \alpha) + C^{BB}_{\ell} \cos^2(2\Delta \alpha) .
\end{equation}
Similarly, this rotation generates a spurious $EB$ cross-spectrum given by:
\begin{equation}
\label{eq:eb-obs}
    C^{EB,\text{obs}}_{\ell} =  \frac{1}{2}(C^{EE}_{\ell} - C^{BB}_{\ell}) \sin(4\Delta \alpha) .
\end{equation}
In the standard model, the intrinsic cosmological $EB$ signal is expected to vanish. Consequently, several standard pipelines exploit the observed $C^{EB,\text{obs}}_{\ell}$ to self-calibrate the instrument and solve for $\Delta \alpha$. It has been demonstrated that this method can be applied even when the $EB$ intrinsic correlation is non zero, relying on prior knowledge of the theoretical $C^{EE}_{\ell} - C^{BB}_{\ell}$ difference \cite{Minami:2020b}. Nevertheless, because this systematic signature perfectly mimics the signal of cosmic birefringence, applying this self-calibration requires extreme caution. Disentangling these effects is further complicated by the interplay between instrumental mis-calibration and foreground cleaning \cite{Diego-Palazuelos:2023}. Additionally, current models of Galactic polarized emission do not fully capture the intrinsic $EB$ correlation observed in \textit{Planck} data \cite{Ritacco:2023, Vacher:2023b, Guillet:2026}. Thus, in order to break the degeneracy between instrumental leakage, complex foregrounds, and cosmological signals, a robust absolute calibration strategy is essential. The COSmological Microwave Observations Calibrator (COSMOCal) project addresses this need by providing an artificial polarized source dedicated to calibrating large-aperture ground-based telescopes. This work is organized as follows: Section \ref{sec:science} outlines the scientific motivations and the specific deliverables of the COSMOCal project, while Section \ref{sec:cosmocal} presents its current design for space. Section \ref{sec:foregrounds} details the impact of foreground complexity on the reconstruction of the cosmological signal. Finally, Section \ref{sec:concl} summarizes our conclusions and discusses future developments.

\section{SCIENTIFIC MOTIVATION}
\label{sec:science}

Reaching the target sensitivity of next-generation experiments ($r < 0.01$) requires tight control of instrumental systematics. Moreover, the large dynamic range of sky signals makes accurate calibration essential to avoid systematic biases that can obscure the cosmological signal. As already mentioned, mis-calibration of the instrumental polarization angle mixes polarization modes, causing $E$-modes to leak into $B$-modes. Similarly, imperfections in the polarization beam profiles can create spurious signals if left uncorrected \cite{Shimon:2008, Bennett:2013}, as recently observed by the SPT-3G and ACT collaborations \cite{Ge:2025, Louis:2025}. To face these effects, absolute calibration of the polarization angle must reach $0.1^{\circ}$ accuracy at $3\sigma$ level. This is necessary to verify current hints of a cosmic birefringence angle of $\beta \sim 0.2^{\circ}$ \cite{Diego-Palazuelos:2025} with a signal-to-noise ratio (SNR) of 6, enabling a robust and independent detection. Achieving sub-degree absolute accuracy remains challenging due to limitations in current calibration methods, which typically rely on astrophysical sources or self-calibration. The standard microwave celestial calibrator is the Crab Nebula (Tau A) \cite{Aumont:2018}, but its polarized emission is complex, frequency-dependent, and diffuse. The polarization angle of Tau A is known to only about $\sim1^{\circ}$ accuracy, varying by frequency band \cite{Ritacco:2018}, which does not meet the required accuracy for next-generation CMB observatories. As already mentioned, experiments often use $EB$ nulling self-calibration, based on the assumption that the cosmological $EB$ cross-correlation power spectrum vanishes \cite{Keating:2013}. While this minimizes instrumental rotation to detect primordial $B$-modes, it also absorbs any cosmic birefringence signal, hiding signs of parity violation or axion-like dark matter. This calibration challenge is further complicated by diffuse thermal dust and synchrotron emissions, which dominate the microwave sky outside a narrow cosmological window \cite{PlanckCollaboration:2020XI-dust}. 

Foreground polarization physics is highly complex, exhibiting spatial and frequency-dependent variations of the polarization angle, alongside intrinsic $EB$ correlations sourced by polarized mixing in the interstellar medium (ISM) \cite{Vacher:2023b, Ritacco:2023, Guillet:2026}. Unmodeled foreground features can then appear as either instrumental misalignments or cosmic birefringence. Characterizing this background is thus simultaneously a need for cosmological analyses and a primary objective for Galactic astrophysics. The planned SO Large Aperture Telescope (SO-LAT) survey will map 40\% of the sky across six frequency bands with arcminute angular resolution \cite{Abitbol:2025}, delivering a powerful dataset for ISM physics. This survey will enable high angular resolution follow-up observations with leading European facilities, specifically the IRAM 30-meter Telescope and the Sardinia Radio Telescope (SRT). This synergy highlights the need for consistent absolute polarization calibration across astrophysical observatories and CMB experiments to effectively combine and exploit their datasets. To overcome the limitations of the aforementioned calibration methods, artificial absolute calibrators can be deployed \cite{Johnson:2015}. Ground- or drone-based solutions, such as the POLOCALC project used for the SO Small Aperture Telescopes (SO-SATs) \cite{Nati:2017, Coppi:2025}, are successful for small apertures but become insufficient for large telescopes due to the far-field condition extending to several kilometers. The COSMOCal project addresses these challenges by deploying a stable, completely linearly polarized artificial microwave source in a geostationary orbit (GEO), directly observable by a designated set of pilot observatories: the SO-LAT (Atacama Desert), the IRAM 30m Telescope (Spain), and the SRT (Italy). Within the SO framework, this space-based calibration optimizes the complementary roles of its sub-systems: while the three $0.4$-meter SO-SATs survey $10\%$ of the sky to constrain primordial $B$-modes, the $6$-meter SO-LAT provides the high-resolution CMB lensing maps required for delensing, while simultaneously acting as the primary instrument to probe cosmic birefringence. Once calibrated via COSMOCal, these observatories will produce reference maps of the CMB and polarized astrophysical sources. These datasets will serve as secondary standards to cross-calibrate facilities unable to observe the COSMOCal source directly, including the South Pole Telescope (SPT) and the \textit{LiteBIRD} satellite, thereby providing a powerful tool for the broader microwave and cosmological community. 

\section{THE COSMOC\MakeLowercase{al} PROJECT: INSTRUMENT OVERVIEW}
\label{sec:cosmocal}
\begin{figure}
    \centering
    \includegraphics[width=0.7\linewidth]{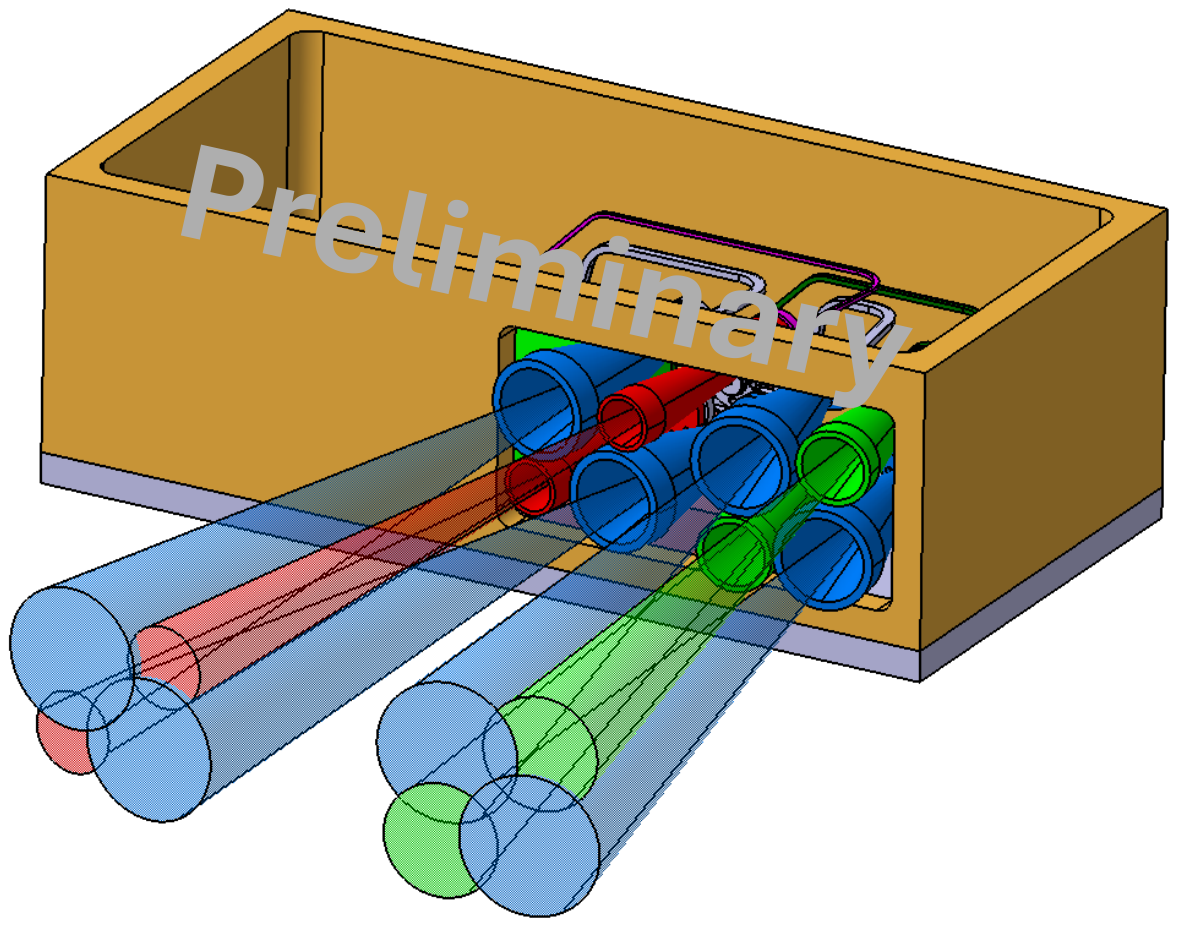}
    \caption{Schematic of the COSMOCal payload. Four fixed pairs of horns target observation sites in Europe and Chile: two pairs operating at 90 and 150~GHz are oriented toward Chile, one 270~GHz pair targets the IRAM 30m telescope, and one 90~GHz pair targets the SRT.}
    \label{fig:design}
\end{figure}

Started in 2022, the COSMOCal project is currently in its development phase under the monitoring of the Centre National d'\'Etudes Spatiales (CNES). This follows a successful laboratory characterization \cite{Ritacco:2024} and an initial testing campaign at the IRAM 30m telescope of a 260~GHz prototype \cite{Savorgnano:2026}. The design is currently being optimized, in order to make it suitable for space. A schematic of the space instrument layout is presented in Fig.~\ref{fig:design}. The payload features eight fixed horns aligned with either SO or the European observatories. Specifically, two pairs operating at 90 and 150~GHz are pointed toward Chile, one pair at 270~GHz targets the IRAM 30m telescope, and one pair at 90~GHz targets the SRT. The 90 and 150~GHz windows towards Chile offer the highest signal to noise ratio for ground-based CMB measurements, the 270~GHz and 90 GHz channels towards Europe will allow to investigate respectively dust polarization at IRAM and  magnetic fields in extragalactic sources at SRT. The horns operate in pairs to emit fully linearly polarized signals oriented at a 45$^{\circ}$ relative offset toward each observatory. Using two horns per frequency allows to have two independent orientations of the output polarization. Additionally, the absolute orientation of the instrument relative to the Earth-centered coordinate system depends directly on the attitude and orientation of the host satellite platform. The COSMOCal source is indeed designed as a guest payload on a geostationary satellite operated by Eutelsat, with a scheduled launch in 2030. The GEO offers a fixed and stable position in the sky, greatly simplifying the tracking requirements for large-aperture telescopes. The source can therefore be observed over long integration times at constant elevation, ensuring stable background conditions. In particular, the host satellite will be positioned at 7.3° West longitude above the Atlantic Ocean between Africa and Brazil. From this orbital location, the source will be observed at an elevation of approximately 40° by the IRAM and SRT telescopes, and around 20° by observatories located in northern Chile, including the SO site. \sil{The emitted signal will be further modulated to disentangle atmospheric background, and switched ON-OFF depending on the observing strategy of the pilot observatories.} The formal CNES Phase A review is expected in early 2027 to evaluate and validate the 2030 launch plan. At the same time, alternative options for a later launch are under evaluation for a backup plan, should the baseline Eutelsat launch be unfeasible.

\section{Impact of Foreground Complexity on Calibration Strategies}
\label{sec:foregrounds}

\subsection{Complexity of dust emission}
\label{sec:dust}
The main polarized contaminants of the microwave sky are the synchrotron and the thermal dust emission. In this work, we focus on analyzing the impact of Galactic dust complexity, which dominates over the CMB signal at high frequencies \cite{Pan-ExperimentGalacticScienceGroup:2025}. This component is produced by the thermal emission of interstellar dust grains aligned with the Galactic magnetic field, and its Spectral Energy Distribution (SED) is canonically modelled as a Modified Black Body (MBB) \cite{Desert:1990}:
\begin{equation}
\label{eq:mbb_canonical}
I(\nu, \hat{n}) = A(\hat{n}) \left(\frac{\nu}{\nu_0}\right)^{\beta(\hat{n})} \frac{B_{\nu}(T(\hat{n}))}{B_{\nu_0}(T(\hat{n}))} ,
\end{equation}
where $A(\hat{n})$ represents the dust intensity template at a reference frequency $\nu_0$ (fixed at $353\,\text{GHz}$ in this analysis), while $B_\nu(T)$ denotes the standard Planck blackbody function. In this formulation, physical conditions in the ISM allow both the spectral index $\beta(\hat{n})$ and the dust temperature $T(\hat{n})$ to vary as a function of the line of sight $\hat{n}$ \cite{Tassis:2015}, well reproducing the observations by \textit{Planck}. However, when integrating along the line of sight or within a finite instrumental beam, spatial variations and averaging effects introduce spectral distortions that cause the observed signal to deviate from a pure MBB. Highly sensitive next-generation CMB experiments are expected to be sensitive to these features. To capture these SED distortions in a model-independent framework, a formalism known as \textit{moment expansion} has been developed \cite{Chluba:2017, Vacher:2022}. This approach relies on Taylor expansion of the intensity field in Eq.~\eqref{eq:mbb_canonical} around fixed reference spectral parameters, namely $\beta_0$ and $T_0$. \sil{By performing the expansion, the total observed intensity map can be parametrized through successive spectral moments, $\omega^{\boldsymbol{p}}_{\alpha}$, up to order $\alpha$, with respect to any parameter $\boldsymbol{p}$. For this work, we use this formalism to describe the dust emission, so that $\boldsymbol{p} = \{\beta, T\}$, and the intensity map can be written as:}

\begin{equation}
\label{eq:momexpdust}
I\left(\nu,\hat{n}\right) =
\frac{I_\nu\left(\beta_0,T_0\right)}
     {I_{\nu_0}\left(\beta_0,T_0\right)}
\Bigg[
A\left(\hat{n}\right)
+ \omega_1^\beta\left(\hat{n}\right)
  \ln\!\left(\frac{\nu}{\nu_0}\right)
+ \frac{1}{2}\,
  \omega_2^\beta\left(\hat{n}\right)
  \ln^2\!\left(\frac{\nu}{\nu_0}\right)
+ \omega_1^T\left(\hat{n}\right)
  \left(
    \Theta\left(\nu,T_0\right)
    - \Theta\left(\nu_0,T_0\right)
  \right)
+ \ldots
\Bigg]
\end{equation}
where $\Theta\left(\nu,T\right) = \frac{x}{T} \frac{e^x}{e^x-1}$, being $x = h\nu/kT$. For a comprehensive review of the mathematical framework of the moment expansion and its cosmological applications, we refer the reader to Refs.~\citenum{Vacher:2023a, Vacher:2025}. Recent studies have shown that this formalism is able to reproduce the observed spectral complexity of dust, in particular to characterize discrepancies between the $EE$ and $BB$ power spectra \cite{Vacher:2023b}. A further complexity in dust modelling is the existence of an intrinsic, non-zero $EB$ correlation. In this work we include it through the model presented in Refs.~\citenum{Hervias-Caimapo:2022, Hervias-Caimapo:2025}. This model is composed of dust filaments imperfectly aligned with the magnetic field, and it reproduces the main features observed in the $Planck$ 353~GHz data. To evaluate the joint impact of these effects, we use the spatial maps generated by this filamentary model as the amplitude templates, $A(\hat{n})$, within the moment expansion framework of Eq.~\eqref{eq:momexpdust}. This scaling law is then used to extrapolate the dust templates across all observational frequencies. Specifically, we evaluate the performance of the pipeline across the standard \texttt{d1} and \texttt{d10} sky models provided by the PySM \cite{Thorne:2017} package, and we compare them against modified realizations that include the intrinsic dust $EB$ templates, denoted hereafter as \texttt{d1EB} and \texttt{d10EB}, respectively. \sil{A complete review of the currently available foregrounds models is provided in Ref.~\citenum{Pan-ExperimentGalacticScienceGroup:2025}. We present here some more details on the ones employed in this work:}

\begin{itemize}
    \item \texttt{d1}: this model is built with polarization $Q$ and $U$ map templates from the \textit{Planck} 2015 data at 353~GHz and extrapolated at all frequencies using a MBB, considering the spatially varying $T$ and $\beta$ obtained from the \textit{Planck} data using the Commander algorithm \cite{PlanckCollaboration:2016};
    
    \item \texttt{d10}: this model is built from the same \textit{Planck} maps but extrapolated at all frequencies using a MBB with varying $\beta$ and $T$ between pixels across the sky, based on templates from the generalized needlet ILC analysis of \textit{Planck} data \cite{Remazeilles:2011};

    \item \texttt{d1EB}: this model is built by computing the moment maps from the \texttt{d1} model and including the filamentary dust template described in Ref.~\citenum{Hervias-Caimapo:2025} in the $A(\hat{n})$ term of Eq.~\eqref{eq:momexpdust};
    
    \item \texttt{d10EB}: this model is built similarly to the previous case, but computing the moment maps from the \texttt{d10} model.

\end{itemize}
\sil{In a fully realistic scenario, further complexities must be expected. In fact, the $EB$ signal can itself exhibit spectral distortions, driven again by polarization mixing. While we leave the detailed study of this scenario to future work, we note that the formalism presented in this paper can naturally accommodate both effects.}

\subsection{Impact on calibration of CMB satellites} 
The absolute polarization calibration delivered by COSMOCal establishes a critical benchmark for next-generation CMB experiments. Once the SO-LAT is calibrated to the targeted $0.1^{\circ}$ accuracy using the COSMOCal reference, its high-fidelity CMB polarization data can be utilized to cross-calibrate smaller telescopes, \sil{a procedure that can be implemented either directly through map-based comparisons or via angular power spectra. While the development and effectiveness of a direct map-based cross-calibration framework is deferred to a future dedicated work, here we focus on the spectrum-based approach, where the SO-LAT spectra can provide a precise measurement, or a stringent upper bound, on the cosmic birefringence angle $\beta$, thereby removing the dominant systematic degeneracy that currently limits the interpretation of the observed $EB$ cross-power spectrum.} From a pipeline perspective, this transforms the standard instrumental self-calibration paradigm: the conventional $EB$ nulling optimization is effectively transposed into an $EB$ matching procedure, following the SO-LAT constraint. Under this framework, any residual $EB$ power observed after the calibration cannot be attributed to a global rotation degeneracy; instead, it must originate either from unmodeled instrumental systematics or from Galactic foreground leakage. Following this spirit, this work quantifies the residual $EB$ contamination expected from complex dust emissions within a \textit{LiteBIRD}-like satellite configuration \cite{LiteBIRDCollaboration:2023}. The analysis adopts a methodology consistent with the current calibration strategy outlined for the mission \cite{Krachmalnicoff:2022}. However, while prior studies restricted their validation to simple foreground scenarios (in particular, the \texttt{d0s0} model, where no spectral variations of the SED are expected for both dust and synchrotron), the study presented here extends the analysis to more complex dust structures, incorporating the dust model defined in Eq.~\eqref{eq:momexpdust}, and investigating the effectiveness of blind component separation methods under realistic conditions in reconstructing the $EB$ cross-spectra.

\subsection{Methods}
To evaluate the impact of complex dust emission under realistic observational conditions, we simulate data expected from a \textit{LiteBIRD}-like satellite configuration. This space-borne experiment is designed to observe the microwave sky across 15 frequency channels, achieving an overall polarization sensitivity of $2.2\,\mu\text{K-arcmin}$. For each channel, we adopt the specific angular resolutions and sensitivity parameters reported in Table~13 of Ref.~\citenum{LiteBIRDCollaboration:2023}. \sil{We generate 100 independent realizations of the microwave sky at a pixel resolution of $\sim 7~\mathrm{arcmin} \, (N_{\mathrm{side}} = 512)$. Each simulation includes an independent realization of the CMB signal and instrumental white noise, to which a foreground contribution is added. The CMB components are generated from theoretical angular power spectra computed using the \texttt{CAMB} package \cite{Lewis:2000}, assuming the best-fit cosmological parameters from the Planck 2018 results \cite{PlanckCollaboration:2020I-cosmology}.} No primordial tensor modes are included (i.e., $r = 0$). This framework yields 100 simulated multi-frequency maps for each of the evaluated dust models: the baseline \texttt{PySM} models (\texttt{d1} and \texttt{d10}) and their modified counterparts incorporating intrinsic dust $EB$ correlations (\texttt{d1EB} and \texttt{d10EB}). 
Furthermore, we include the \texttt{PySM} \texttt{s1} model to account for Galactic synchrotron emission \cite{Delabrouille:2013}. No additional complexity or intrinsic $EB$ correlations are included for the synchrotron, as no such signal has been detected to date \cite{Martire:2022}; moreover, Galactic dust is expected to be the dominant foreground contaminant for \textit{LiteBIRD}. All frequency maps are initially co-added and smoothed according to their nominal channel beams, so that in each frequency channel $\nu$, the observed signal can be expressed in terms of the Stokes parameters as:
\begin{equation}
    \mathbf{S}_{\nu} = 
    \mathbf{B}(\theta_{\nu})
    \begin{bmatrix} 
        \mathbf{A}_{\text{CMB}} & \mathbf{A}_{\text{dust}}^{\nu} & \mathbf{A}_{\text{synch}}^{\nu} 
    \end{bmatrix}
    \begin{bmatrix} 
        \mathbf{S_{\text{CMB}}} \\ 
        \mathbf{S_{\text{dust}}} \\
        \mathbf{S_{\text{synch}}}
    \end{bmatrix} 
    + \mathbf{n}_{\nu} \, ,
\end{equation}
where $\mathbf{n}_{\nu}$ is the white noise contribution. Subsequently, all maps are brought to a common angular resolution of $30.0\,\text{arcmin}$ prior to component separation. Similarly, we generate a twin set of simulations featuring the exact same 100 CMB and noise realizations, but incorporating the presence of polarization angle mis-calibration. In this scenario, a rotation operation is applied directly to the input maps, modifying the observed signal as follows:
\begin{equation}
    \mathbf{S}^{\text{rot}}_{\nu} = 
    \mathbf{R}(\alpha_{\nu}) \mathbf{B}(\theta_{\nu})
    \begin{bmatrix} 
        \mathbf{A}_{\text{CMB}} & \mathbf{A}_{\text{dust}}^{\nu} & \mathbf{A}_{\text{synch}}^{\nu} 
    \end{bmatrix}
    \begin{bmatrix} 
        \mathbf{S_{\text{CMB}}} \\ 
        \mathbf{S_{\text{dust}}} \\
        \mathbf{S_{\text{synch}}}
    \end{bmatrix} 
    + \mathbf{n}_{\nu} \, .
\end{equation}
In this analysis, we consider an overall mis-calibration angle of $\Delta \alpha = 0.1^{\circ}$, corresponding to the COSMOCal target accuracy requirement.

To isolate the CMB signal from Galactic contaminants without relying on parametric models for the foreground SEDs, we employ a blind component-separation method. This class of algorithms requires no prior assumptions on foreground physics, aside from treating the CMB as a perfect blackbody. The baseline architecture of these methods is the Internal Linear Combination (ILC) technique \cite{Bennett:2003b, Eriksen:2004}. Considering $k$ observed frequency maps expressed in thermodynamic temperature units, the total signal is modelled as $T(\nu_k) = T_{\mathrm{CMB}} + T_{\mathrm{res}}(\nu_k)$, where $T_{\mathrm{res}}(\nu_k)$ accounts for the combined contribution of foregrounds and instrumental noise, assumed to be statistically independent of the CMB. The ILC forms a weighted linear combination of the multi-frequency channels:
\begin{equation}
T = \sum_{k} w_k\, T(\nu_k) ,
\end{equation}
subject to the constraint $\sum_k w_k = 1$ to preserve the CMB blackbody spectrum. The weights $w_k$ are determined by minimizing the total variance of the reconstructed map. While a global ILC can perform poorly over large scales due to spatial and spectral variations of Galactic emissions, localization techniques significantly improve performance. Needlet ILC (NILC) implements a multi-scale decomposition using needlets, providing simultaneous localization in both pixel and harmonic space \cite{Basak:2013}. To capture further complexities in polarized Galactic $B$-mode emission, the partitioning can be optimized via clustering algorithms. This defines the Multi-Clustering NILC (MC-NILC) framework \cite{Carones:2023}, which groups sky patches sharing similar spectral properties and merges the localized reconstructions into a single, full-sky map. In this work, we implement a blind component separation through the publicly available package \texttt{BROOM} \cite{Carones:2026b}, validating our framework against recent release benchmarks \cite{Carones:2026a}. To optimize the extraction of the $B$-mode signal, we adopt a hybrid approach hereafter defined as realistic MC-NILC (\text{rMC-NILC}). Under this configuration, MC-NILC is applied exclusively to the first needlet band using data-driven construction of clusters via the Generalized ILC (GILC) method. This restriction ensures that the complex spatial partitioning is confined to the largest angular scales, where the signal-to-noise ratio allows for an accurate patch reconstruction. For all the subsequent needlet bands, a standard NILC is applied. Since this algorithm is currently well optimized for $B$-mode reconstruction, the $E$-mode maps are reconstructed through standard NILC. The independent $E$ and $B$ products are ultimately combined using the \texttt{\_combine\_products} routine within \texttt{BROOM}. Finally, the angular power spectra are computed using the \texttt{compute\_spectra} function in \texttt{BROOM}, which implements the pseudo-$C_\ell$ MASTER formalism \cite{Hivon:2002} via the \texttt{NaMaster} package \cite{Alonso:2019}. All power spectra are evaluated using a 60~\% Galactic mask derived from the \textit{Planck} 2018 data\footnote{\url{https://irsa.ipac.caltech.edu/data/Planck/release_2/ancillary-data/previews/HFI_Mask_GalPlane-apo5_2048_R2.00/index.html}} and a binning scheme with $\Delta \ell  = 5 $. The results presented in Sec.~\ref{sec:results} represent the ensemble average over the 100 independent realizations processed through the pipeline. 

\subsection{Results}
\label{sec:results}
The pipeline is first validated against the low-complexity \texttt{d1} Galactic dust model from \texttt{PySM}, successfully recovering unbiased $BB$ spectra, in agreement with Fig.~7 of Ref.~\citenum{Carones:2026b}. We expect an ILC to be generally capable of absorbing frequency-dependent systematic effects. This mitigation capability of blind methods, due to their tendency to absorb instrumental contamination into the foreground weights, is consistent with previous comparisons with parametric approaches \cite{Krachmalnicoff:2022}. However, in the presence of a global, frequency-independent rotation, as in the case considered here, residual contamination is still expected after component separation. We quantify the impact of this contamination on the reconstruction of the cross-angular power spectra by evaluating them on mis-calibrated maps after component separation.

\begin{figure}[htbp]
    \centering
    \includegraphics[width=0.8\linewidth]{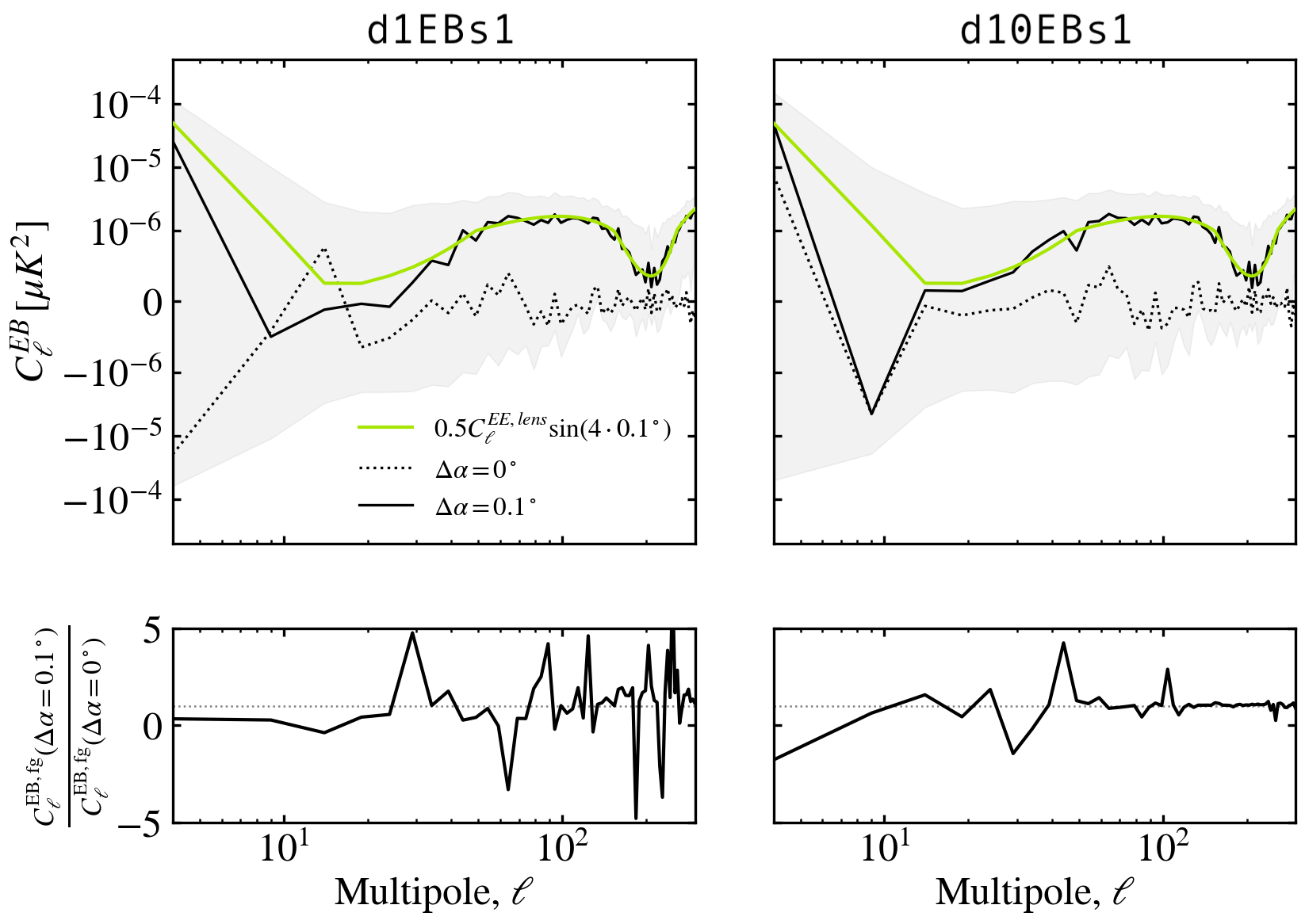}
    \caption{\textit{Top}: Recovered $EB$ angular power spectra after rMC-NILC for the $\mathtt{d1EBs1}$ (left) and $\mathtt{d10EBs1}$ (right) dust models averaged over 100 simulations. Black curves show the total output spectra obtained without instrumental rotation ($\Delta\alpha = 0^\circ$, dotted) and with an injected mis-calibration angle ($\Delta\alpha = 0.1^\circ$, solid). The green line indicates the theoretical $EB$ spectrum expected from a pure $0.1^\circ$ cosmological rotation, computed as $0.5 C_{\ell}^{EE,\text{lens}} \sin(4\Delta \alpha)$. \textit{Bottom}: Ratio between the residual dust power spectrum in the rotated sky configuration ($\Delta \alpha = 0.1^\circ$) and the same component in the un-rotated scenario ($\Delta \alpha = 0^\circ$) averaged over 100 simulations. The horizontal dashed line indicates the unity.}
    \label{fig:ClEB_d1_d10_EB_rotated_comparison}
\end{figure}

The upper panels of Fig.~\ref{fig:ClEB_d1_d10_EB_rotated_comparison} show the $EB$ spectra for the $\mathtt{d1EBs1}$ (left panel) and $\mathtt{d10EBs1}$ (right panel) models, comparing the un-rotated ($\Delta \alpha=0^{\circ}$, dotted lines) and the rotated scenario (solid lines), assuming a polarization angle mis-calibration of $\Delta \alpha = 0.1^{\circ}$. The reconstruction of the $EB$ cross-spectrum remains significantly biased in the presence of a mis-calibration. We also show with the green lines the theoretical $EB$ spectrum arising from a pure cosmological rotation (for example due to cosmic birefringence) of $0.1^{\circ}$. According to Eq.~\eqref{eq:eb-obs}, in the case lensing only $E$ modes are present, this signal corresponds to $C^{EB,\text{obs}}_{\ell} = 1/2 \, C^{EE,\mathrm{lens}}_{\ell} \sin(4\Delta \alpha)$. While the primary quantities evaluated in our analysis are the total recovered power spectra, the linearity of the ILC also allows us to apply the optimized weights directly to individual simulated components (foregrounds and noise) to inspect their specific contributions. For example, in the bottom panels of Fig.~\ref{fig:ClEB_d1_d10_EB_rotated_comparison} we show the ratio of the residual foreground spectra after component separation between the rotated and the un-rotated case. We observe that the residual foreground power spectra exhibit a slightly larger residual at low multipoles, particularly for the most complex scenario. 

\begin{figure}[htbp]
    \centering
    \includegraphics[width=0.8\linewidth]{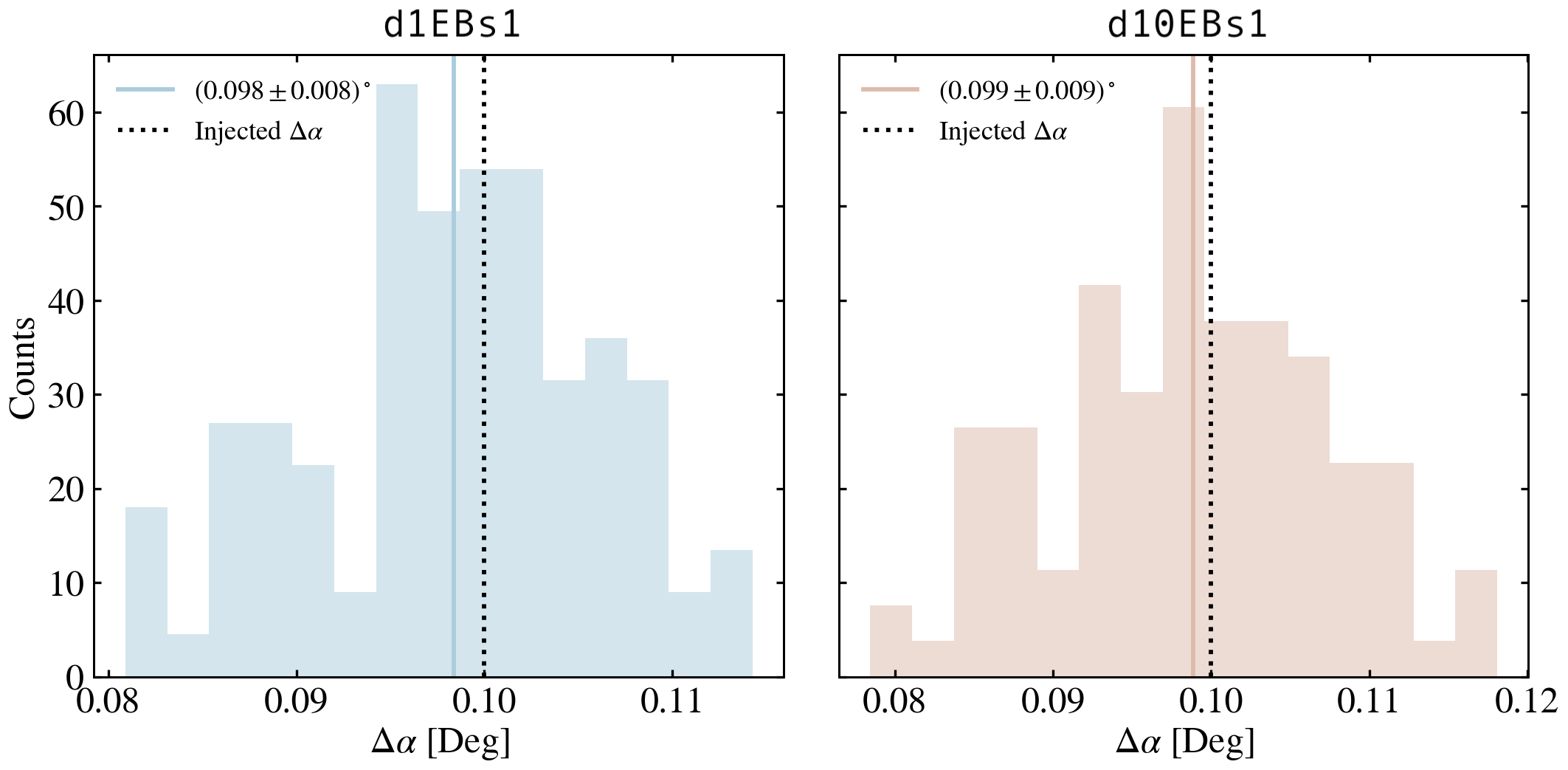}
    \caption{Posterior distribution of the reconstructed mis-calibration angle $\Delta \alpha$ across 100 independent realizations for the $\mathtt{d1EBs1}$ (left) and $\mathtt{d10EBs1}$ (right) dust models. In both panels, the black dashed line indicates the injected nominal value ($\Delta \alpha = 0.1^\circ$), while the coloured solid line represents the ensemble mean.}
    \label{fig:hist_alpha}
\end{figure}

We can  use the information on the reconstructed $EB$ spectra to estimate the mis-calibration angle. To do so, we employ a standard $\chi^2$ approach to minimize the residual $EB$ expectation value. For each multipole bin, the rotated estimator is thus defined as:
\begin{equation}
\Tilde{C}_{\ell}^{EB}(\Delta \alpha) = C_{\ell}^{EB,\text{obs}} \cos(4\Delta \alpha) - \frac{1}{2} \left( C_{\ell}^{EE,\text{obs}} - C_{\ell}^{BB,\text{obs}} \right) \sin(4\Delta \alpha) \, ,
\end{equation}
where all spectra represent the reconstructed quantities after component separation. For each independent realization, we consider the binned $\chi^2$ function:
\begin{equation}
\chi^2(\Delta \alpha) = \mathbf{\Tilde{C}}^{EB}(\Delta \alpha)^{T} \, \mathbf{\Sigma}_{EB}^{-1} \, \mathbf{\Tilde{C}}^{EB}(\Delta \alpha) \, ,
\end{equation}
where $\mathbf{\Sigma}_{EB}$ is the binned $EB$ covariance matrix estimated directly from the simulation ensemble. The best-fit angle $\alpha_{\text{best}}$ is determined by minimizing this profile, while its associated $1\sigma$ statistical uncertainty is derived from the standard $\Delta\chi^2 = 1$ threshold. The results of this analysis are summarized in Fig.~\ref{fig:hist_alpha}. In both sky configurations, the injection angle of $\Delta \alpha = 0.1^\circ$ is successfully recovered with high accuracy, yielding an ensemble average of $\Delta \alpha_{\text{best}} = 0.098^\circ$ for the $\mathtt{d1EBs1}$ case and $\Delta \alpha_{\text{best}} = 0.099^\circ$ for the $\mathtt{d10EBs1}$ case. The $1\sigma$ relative statistical uncertainty is found to be $7.7\%$ for $\mathtt{d1EBs1}$ and $8.7\%$ for $\mathtt{d10EBs1}$. To further test the impact of large scale residual foreground features, we performed a dedicated check by restricting the minimization exclusively to the low-$\ell$ region ($\ell < 40$). In this scenario, the reconstructed calibration angle for both models exhibits a systematic bias, significantly deviating from the injected value. In particular, we find $(0.052 \pm 0.14)^\circ$ and $(0.10 \pm 0.13)^\circ$ for $\mathtt{d1EBs1}$ and $\mathtt{d10EBs1}$, respectively. Conversely, when we expand the analysis to incorporate the full available multipole range, specifically extending the fit up to $\ell_{\text{max}} = 2N_{\text{side}} = 1024$, the statistical weight of the intermediate and high multipoles effectively stabilizes the likelihood. In conclusion, this analysis demonstrates that by employing a more refined and physically grounded framework, we are able to robustly characterize the mis-calibration angle after component separation with a significant improvement with respect to previous studies~\cite{Krachmalnicoff:2022}. 

\section{CONCLUSIONS}
\label{sec:concl}
Future CMB polarization experiments require absolute calibration of instrumental polarization angles to sub-degree precision. Traditional self-calibration, based on the assumption of a vanishing intrinsic $EB$ cross-correlation, could fail to disentangle a systematic mis-calibration from cosmological cosmic birefringence or astrophysical emissions. On the other hand, astrophysical polarized sources are not accurate enough for next-generation sensitivities. In this scenario, the COSMOCal project plays a crucial role by providing a unique, space-based absolute polarization calibrator. By deploying a stable, completely linearly polarized artificial microwave source in a geostationary orbit, COSMOCal decouples the instrumental calibration from the astrophysical sky, delivering a reference that helps break these degeneracies. This source will be directly observable by a designated set of pilot observatories: the SO-LAT, the IRAM 30m Telescope, and SRT. Once calibrated via COSMOCal, these observatories will produce reference maps of the CMB and polarized astrophysical sources. These datasets will serve as secondary standards to cross-calibrate facilities unable to observe the COSMOCal source directly, including SPT and the \textit{LiteBIRD} satellite, thereby providing a powerful tool for the broader microwave and cosmological community. In this work, we investigated the implications of this calibration strategy for the case of a \textit{LiteBIRD}-like experiment. Unlike standard $EB$-nulling techniques, the framework considered here leverages the cosmic birefringence angle measurements by SO-LAT after its absolute calibration through COSMOCal. Consequently, any residual $EB$ signal observed after calibration cannot be attributed to a cosmological global rotation and must instead originated from instrumental systematic effects or astrophysical contamination. Under this framework, we showed that a non-negligible residual $EB$ signal persists after component separation when realistic and complex foreground models are considered. This systematic leakage was already observed in previous studies \cite{Krachmalnicoff:2022}; here, we extend the analysis by highlighting the importance of incorporating realistic foreground complexity when assessing calibration performance. We also presented a formalism capable of describing a non-zero dust $EB$ correlation, providing a general framework for future applications in which the $EB$ signal itself may exhibit spectral distortions. Within this context, we explored the feasibility of reconstructing the mis-calibration angle directly from the recovered $EB$ cross-spectra through a $\chi^2$ minimization procedure. We showed that the injected mis-calibration angle can be recovered with significantly higher accuracy than in previous studies. Overall, our results demonstrate that characterizing the complexity of Galactic dust is not only essential for cosmological analyses, but also for the calibration strategies of future CMB polarization experiments. Future analyses will naturally build upon this framework to quantify the bias induced on both the cosmic birefringence angle $\beta$ and the tensor-to-scalar ratio $r$ in the presence of complex sky models.

\acknowledgments 
We thank Carlos Herv\'{i}as-Caimapo for providing the data of the dust filamentary model used in this work. SM acknowledges support from the Labex FOCUS.

\bibliography{report} 

@INPROCEEDINGS{Savorgnano:2026,
       author = {{Savorgnano, S. et al.}},
        title = "{The COSMOCal project: proof of concept at the IRAM 30m telescope with NIKA2}",
         year = 2026,
       series = {Society of Photo-Optical Instrumentation Engineers (SPIE) Conference Series,},
       note = {[In prep.]}
}

@ARTICLE{Remazeilles:2011,
       author = {{Remazeilles}, Mathieu and {Delabrouille}, Jacques and {Cardoso}, Jean-Fran{\c{c}}ois},
        title = "{Foreground component separation with generalized Internal Linear Combination}",
      journal = {MNRAS},
         year = 2011,
        month = nov,
       volume = {418},
       number = {1},
        pages = {467-476},
          doi = {10.1111/j.1365-2966.2011.19497.x},
archivePrefix = {arXiv},
       eprint = {1103.1166},
 primaryClass = {astro-ph.CO},
       adsurl = {https://ui.adsabs.harvard.edu/abs/2011MNRAS.418..467R}
}

@ARTICLE{Bennett:2003b,
       author = {{Bennett}, C.~L. and {Hill}, R.~S. and {Hinshaw}, G. and {Nolta}, M.~R. and {Odegard}, N. and {Page}, L. and {Spergel}, D.~N. and {Weiland}, J.~L. and {Wright}, E.~L. and {Halpern}, M. and {Jarosik}, N. and {Kogut}, A. and {Limon}, M. and {Meyer}, S.~S. and {Tucker}, G.~S. and {Wollack}, E.},
        title = "{First-Year Wilkinson Microwave Anisotropy Probe (WMAP) Observations: Foreground Emission}",
      journal = {ApJS},
         year = 2003,
        month = sep,
       volume = {148},
       number = {1},
        pages = {97-117},
          doi = {10.1086/377252},
archivePrefix = {arXiv},
       eprint = {astro-ph/0302208},
 primaryClass = {astro-ph},
       adsurl = {https://ui.adsabs.harvard.edu/abs/2003ApJS..148...97B}
}

@ARTICLE{Basak:2013,
       author = {{Basak}, Soumen and {Delabrouille}, Jacques},
        title = "{A needlet ILC analysis of WMAP 9-year polarization data: CMB polarization power spectra}",
      journal = {MNRAS},
         year = 2013,
        month = oct,
       volume = {435},
       number = {1},
        pages = {18-29},
          doi = {10.1093/mnras/stt1158},
archivePrefix = {arXiv},
       eprint = {1204.0292},
 primaryClass = {astro-ph.CO},
       adsurl = {https://ui.adsabs.harvard.edu/abs/2013MNRAS.435...18B}
}

@ARTICLE{Eriksen:2004,
       author = {{Eriksen}, H.~K. and {Banday}, A.~J. and {G{\'o}rski}, K.~M. and {Lilje}, P.~B.},
        title = "{On Foreground Removal from the Wilkinson Microwave Anisotropy Probe Data by an Internal Linear Combination Method: Limitations and Implications}",
      journal = {ApJ},
         year = 2004,
        month = sep,
       volume = {612},
       number = {2},
        pages = {633-646},
          doi = {10.1086/422807},
archivePrefix = {arXiv},
       eprint = {astro-ph/0403098},
 primaryClass = {astro-ph},
       adsurl = {https://ui.adsabs.harvard.edu/abs/2004ApJ...612..633E}
}

@ARTICLE{Bennett:2013,
       author = {{Bennett}, C.~L. and {Larson}, D. and {Weiland}, J.~L. and {Jarosik}, N. and {Hinshaw}, G. and {Odegard}, N. and {Smith}, K.~M. and {Hill}, R.~S. and {Gold}, B. and {Halpern}, M. and {Komatsu}, E. and {Nolta}, M.~R. and {Page}, L. and {Spergel}, D.~N. and {Wollack}, E. and {Dunkley}, J. and {Kogut}, A. and {Limon}, M. and {Meyer}, S.~S. and {Tucker}, G.~S. and {Wright}, E.~L.},
        title = "{Nine-year Wilkinson Microwave Anisotropy Probe (WMAP) Observations: Final Maps and Results}",
      journal = {ApJS},
         year = 2013,
        month = oct,
       volume = {208},
       number = {2},
          eid = {20},
        pages = {20},
          doi = {10.1088/0067-0049/208/2/20},
archivePrefix = {arXiv},
       eprint = {1212.5225},
 primaryClass = {astro-ph.CO},
       adsurl = {https://ui.adsabs.harvard.edu/abs/2013ApJS..208...20B}
}

@ARTICLE{Pan-ExperimentGalacticScienceGroup:2025,
       author = {{Pan-Experiment Galactic Science Group} and {Borrill}, Julian and {Clark}, Susan E. and {Delabrouille}, Jacques and {Frolov}, Andrei V. and {Ghosh}, Shamik and {Hensley}, Brandon S. and {Hicks}, Monica D. and {Krachmalnicoff}, Nicoletta and {Lau}, King and {Norton}, Myra M. and {Pryke}, Clement and {Puglisi}, Giuseppe and {Remazeilles}, Mathieu and {Russier}, Elisa and {Thorne}, Benjamin and {Yao}, Jian and {Zonca}, Andrea},
        title = "{Full-sky Models of Galactic Microwave Emission and Polarization at Subarcminute Scales for the Python Sky Model}",
      journal = {ApJ},
         year = 2025,
        month = sep,
       volume = {991},
       number = {1},
          eid = {23},
        pages = {23},
          doi = {10.3847/1538-4357/adf212},
archivePrefix = {arXiv},
       eprint = {2502.20452},
 primaryClass = {astro-ph.CO},
       adsurl = {https://ui.adsabs.harvard.edu/abs/2025ApJ...991...23P}
}

@ARTICLE{Thorne:2017,
       author = {{Thorne}, B. and {Dunkley}, J. and {Alonso}, D. and {N{\ae}ss}, S.},
        title = "{The Python Sky Model: software for simulating the Galactic microwave sky}",
      journal = {MNRAS},
         year = 2017,
        month = aug,
       volume = {469},
       number = {3},
        pages = {2821-2833},
          doi = {10.1093/mnras/stx949},
archivePrefix = {arXiv},
       eprint = {1608.02841},
 primaryClass = {astro-ph.CO},
       adsurl = {https://ui.adsabs.harvard.edu/abs/2017MNRAS.469.2821T}
}

@ARTICLE{Vacher:2025,
       author = {{Vacher}, L. and {Carones}, A. and {Aumont}, J. and {Chluba}, J. and {Krachmalnicoff}, N. and {Ranucci}, C. and {Remazeilles}, M. and {Rizzieri}, A.},
        title = "{How bad could it be? Modelling the 3D complexity of the polarised dust signal using moment expansion}",
      journal = {A\&A},
         year = 2025,
        month = may,
       volume = {697},
          eid = {A212},
        pages = {A212},
          doi = {10.1051/0004-6361/202453066},
archivePrefix = {arXiv},
       eprint = {2411.11649},
 primaryClass = {astro-ph.CO},
       adsurl = {https://ui.adsabs.harvard.edu/abs/2025A&A...697A.212V}
}

@ARTICLE{Vacher:2022,
       author = {{Vacher}, L. and {Aumont}, J. and {Montier}, L. and {Azzoni}, S. and {Boulanger}, F. and {Remazeilles}, M.},
        title = "{Moment expansion of polarized dust SED: A new path towards capturing the CMB B-modes with LiteBIRD}",
      journal = {A\&A},
         year = 2022,
        month = apr,
       volume = {660},
          eid = {A111},
        pages = {A111},
          doi = {10.1051/0004-6361/202142664},
archivePrefix = {arXiv},
       eprint = {2111.07742},
 primaryClass = {astro-ph.CO},
       adsurl = {https://ui.adsabs.harvard.edu/abs/2022A&A...660A.111V}
}

@ARTICLE{Chluba:2017,
       author = {{Chluba}, Jens and {Hill}, James Colin and {Abitbol}, Maximilian H.},
        title = "{Rethinking CMB foregrounds: systematic extension of foreground parametrizations}",
      journal = {MNRAS},
         year = 2017,
        month = nov,
       volume = {472},
       number = {1},
        pages = {1195-1213},
          doi = {10.1093/mnras/stx1982},
archivePrefix = {arXiv},
       eprint = {1701.00274},
 primaryClass = {astro-ph.CO},
       adsurl = {https://ui.adsabs.harvard.edu/abs/2017MNRAS.472.1195C}
}

@ARTICLE{Tassis:2015,
       author = {{Tassis}, K. and {Pavlidou}, V.},
        title = "{Searching for inflationary B modes: can dust emission properties be extrapolated from 350 GHz to 150 GHz?}",
      journal = {MNRAS},
         year = 2015,
        month = jul,
       volume = {451},
        pages = {L90-L94},
          doi = {10.1093/mnrasl/slv077},
archivePrefix = {arXiv},
       eprint = {1410.8136},
 primaryClass = {astro-ph.CO},
       adsurl = {https://ui.adsabs.harvard.edu/abs/2015MNRAS.451L..90T}
}

@ARTICLE{Hivon:2002,
       author = {{Hivon}, Eric and {G{\'o}rski}, Krzysztof M. and {Netterfield}, C. Barth and {Crill}, Brendan P. and {Prunet}, Simon and {Hansen}, Frode},
        title = "{MASTER of the Cosmic Microwave Background Anisotropy Power Spectrum: A Fast Method for Statistical Analysis of Large and Complex Cosmic Microwave Background Data Sets}",
      journal = {ApJ},
         year = 2002,
        month = mar,
       volume = {567},
       number = {1},
        pages = {2-17},
          doi = {10.1086/338126},
archivePrefix = {arXiv},
       eprint = {astro-ph/0105302},
 primaryClass = {astro-ph},
       adsurl = {https://ui.adsabs.harvard.edu/abs/2002ApJ...567....2H}
}

@ARTICLE{Alonso:2019,
       author = {{Alonso}, David and {Sanchez}, Javier and {Slosar}, An{\v{z}}e and {LSST Dark Energy Science Collaboration}},
        title = "{A unified pseudo-C$_{{\ensuremath{\ell}}}$ framework}",
      journal = {MNRAS},
         year = 2019,
        month = apr,
       volume = {484},
       number = {3},
        pages = {4127-4151},
          doi = {10.1093/mnras/stz093},
archivePrefix = {arXiv},
       eprint = {1809.09603},
 primaryClass = {astro-ph.CO},
       adsurl = {https://ui.adsabs.harvard.edu/abs/2019MNRAS.484.4127A}
}

@ARTICLE{Carones:2023,
       author = {{Carones}, Alessandro and {Migliaccio}, Marina and {Puglisi}, Giuseppe and {Baccigalupi}, Carlo and {Marinucci}, Domenico and {Vittorio}, Nicola and {Poletti}, Davide and {LiteBIRD collaboration}},
        title = "{Multiclustering needlet ILC for CMB B-mode component separation}",
      journal = {MNRAS},
         year = 2023,
        month = oct,
       volume = {525},
       number = {2},
        pages = {3117-3135},
          doi = {10.1093/mnras/stad2423},
archivePrefix = {arXiv},
       eprint = {2212.04456},
 primaryClass = {astro-ph.CO},
       adsurl = {https://ui.adsabs.harvard.edu/abs/2023MNRAS.525.3117C}
}

@ARTICLE{Carones:2026a,
       author = {{Carones}, Alessandro},
        title = "{Debiasing cosmological parameters from large-scale foreground contamination in Cosmic Microwave Background data}",
      journal = {JCAP},
         year = 2026,
        month = may,
       volume = {2026},
       number = {5},
          eid = {028},
        pages = {028},
          doi = {10.1088/1475-7516/2026/05/028},
archivePrefix = {arXiv},
       eprint = {2510.20785},
 primaryClass = {astro-ph.CO},
       adsurl = {https://ui.adsabs.harvard.edu/abs/2026JCAP...05..028C}
}

@ARTICLE{Carones:2026b,
       author = {{Carones}, Alessandro and {Jose}, Sijil and {Mustafa}, Aliza and {Krachmalnicoff}, Nicoletta and {Baccigalupi}, Carlo},
        title = "{BROOM: a python package for model-independent analysis of microwave astronomical data}",
      journal = {arXiv},
         year = 2026,
        month = apr,
          eid = {arXiv:2604.14088},
        pages = {arXiv:2604.14088},
          doi = {10.48550/arXiv.2604.14088},
archivePrefix = {arXiv},
       eprint = {2604.14088},
 primaryClass = {astro-ph.CO},
       adsurl = {https://ui.adsabs.harvard.edu/abs/2026arXiv260414088C}
}

@ARTICLE{Guillet:2026,
       author = {{Guillet}, Vincent and {Vacher}, L{\'e}o and {Aumont}, Jonathan and {Boulanger}, Fran{\c{c}}ois and {Ritacco}, Alessia and {Delouis}, Jean-Marc and {Bracco}, Andrea},
        title = "{Variance of dust temperature and spectral index in Planck polarization data using spin-moment expansion}",
      journal = {A\&A},
         year = 2026,
        month = jan,
       volume = {705},
          eid = {A177},
        pages = {A177},
          doi = {10.1051/0004-6361/202556491},
archivePrefix = {arXiv},
       eprint = {2510.18305},
 primaryClass = {astro-ph.GA},
       adsurl = {https://ui.adsabs.harvard.edu/abs/2026A&A...705A.177G}
}

@ARTICLE{Ge:2025,
       author = {{Ge}, F. and {Millea}, M. and {Camphuis}, E. and {Daley}, C. and {Huang}, N. and {Omori}, Y. and {Quan}, W. and {Anderes}, E. and {Anderson}, A.~J. and {Ansarinejad}, B. and {Archipley}, M. and {Balkenhol}, L. and {Benabed}, K. and {Bender}, A.~N. and {Benson}, B.~A. and {Bianchini}, F. and {Bleem}, L.~E. and {Bouchet}, F.~R. and {Bryant}, L. and {Carlstrom}, J.~E. and {Chang}, C.~L. and {Chaubal}, P. and {Chen}, G. and {Chichura}, P.~M. and {Chokshi}, A. and {Chou}, T.-L. and {Coerver}, A. and {Crawford}, T.~M. and {de Haan}, T. and {Dibert}, K.~R. and {Dobbs}, M.~A. and {Doohan}, M. and {Doussot}, A. and {Dutcher}, D. and {Everett}, W. and {Feng}, C. and {Ferguson}, K.~R. and {Fichman}, K. and {Foster}, A. and {Galli}, S. and {Gambrel}, A.~E. and {Gardner}, R.~W. and {Goeckner-Wald}, N. and {Gualtieri}, R. and {Guidi}, F. and {Guns}, S. and {Halverson}, N.~W. and {Hivon}, E. and {Holder}, G.~P. and {Holzapfel}, W.~L. and {Hood}, J.~C. and {Howe}, D. and {Hryciuk}, A. and {K{\'e}ruzor{\'e}}, F. and {Khalife}, A.~R. and {Knox}, L. and {Korman}, M. and {Kornoelje}, K. and {Kuo}, C.-L. and {Lee}, A.~T. and {Levy}, K. and {Lowitz}, A.~E. and {Lu}, C. and {Maniyar}, A. and {Martsen}, E.~S. and {Menanteau}, F. and {Montgomery}, J. and {Nakato}, Y. and {Natoli}, T. and {Noble}, G.~I. and {Pan}, Z. and {Paschos}, P. and {Phadke}, K.~A. and {Pollak}, A.~W. and {Prabhu}, K. and {Rahimi}, M. and {Rahlin}, A. and {Reichardt}, C.~L. and {Riebel}, D. and {Rouble}, M. and {Ruhl}, J.~E. and {Schiappucci}, E. and {Sobrin}, J.~A. and {Stark}, A.~A. and {Stephen}, J. and {Tandoi}, C. and {Thorne}, B. and {Trendafilova}, C. and {Umilta}, C. and {Vieira}, J.~D. and {Vitrier}, A. and {Wan}, Y. and {Whitehorn}, N. and {Wu}, W.~L.~K. and {Young}, M.~R. and {Zebrowski}, J.~A. and {SPT-3G Collaboration}},
        title = "{Cosmology from CMB lensing and delensed EE power spectra using 2019-2020 SPT-3G polarization data}",
      journal = {PhRvD},
         year = 2025,
        month = apr,
       volume = {111},
       number = {8},
          eid = {083534},
        pages = {083534},
          doi = {10.1103/PhysRevD.111.083534},
archivePrefix = {arXiv},
       eprint = {2411.06000},
 primaryClass = {astro-ph.CO},
       adsurl = {https://ui.adsabs.harvard.edu/abs/2025PhRvD.111h3534G}
}

@ARTICLE{Diego-Palazuelos:2025,
       author = {{Diego-Palazuelos}, P. and {Komatsu}, E.},
        title = "{Cosmic Birefringence from the Atacama Cosmology Telescope Data Release 6}",
      journal = {arXiv},
         year = 2025,
        month = sep,
          eid = {arXiv:2509.13654},
        pages = {arXiv:2509.13654},
          doi = {10.48550/arXiv.2509.13654},
archivePrefix = {arXiv},
       eprint = {2509.13654},
 primaryClass = {astro-ph.CO},
       adsurl = {https://ui.adsabs.harvard.edu/abs/2025arXiv250913654D}
}

@ARTICLE{Coppi:2025,
       author = {{Coppi}, Gabriele and {Dachlythra}, Nadia and {Nati}, Federico and {D{\"u}nner-Planella}, Rolando and {Adler}, Alexandre E. and {Errard}, Josquin and {Galitzki}, Nicholas and {Li}, Yunyang and {Petroff}, Matthew A. and {Simon}, Sara M. and {Sang}, Ema Tsang King and {Aguilar}, Amalia Villarrubia and {Wollack}, Edward J. and {Zannoni}, Mario},
        title = "{PROTOCALC, a W-band Polarized Calibrator for Cosmic Microwave Background Telescopes: Application to Simons Observatory and CLASS}",
      journal = {ApJS},
         year = 2025,
        month = jul,
       volume = {279},
       number = {1},
          eid = {30},
        pages = {30},
          doi = {10.3847/1538-4365/adde5f},
archivePrefix = {arXiv},
       eprint = {2502.14473},
 primaryClass = {astro-ph.IM},
       adsurl = {https://ui.adsabs.harvard.edu/abs/2025ApJS..279...30C}
}

@ARTICLE{Louis:2025,
       author = {{Louis}, Thibaut and {La Posta}, Adrien and {Atkins}, Zachary and {Jense}, Hidde T. and {Abril-Cabezas}, Irene and {Addison}, Graeme E. and {Ade}, Peter A.~R. and {Aiola}, Simone and {Alford}, Tommy and {Alonso}, David and {Amiri}, Mandana and {An}, Rui and {Austermann}, Jason E. and {Barbavara}, Eleonora and {Battaglia}, Nicholas and {Battistelli}, Elia Stefano and {Beall}, James A. and {Bean}, Rachel and {Beheshti}, Ali and {Beringue}, Benjamin and {Bhandarkar}, Tanay and {Biermann}, Emily and {Bolliet}, Boris and {Bond}, J. Richard and {Calabrese}, Erminia and {Capalbo}, Valentina and {Carrero}, Felipe and {Chen}, Shi-Fan and {Chesmore}, Grace and {Cho}, Hsiao-mei and {Choi}, Steve K. and {Clark}, Susan E. and {Cothard}, Nicholas F. and {Coughlin}, Kevin and {Coulton}, William and {Crichton}, Devin and {Crowley}, Kevin T. and {Darwish}, Omar and {Devlin}, Mark J. and {Dicker}, Simon and {Duell}, Cody J. and {Duff}, Shannon M. and {Duivenvoorden}, Adriaan J. and {Dunkley}, Jo and {Dunner}, Rolando and {Embil Villagra}, Carmen and {Fankhanel}, Max and {Farren}, Gerrit S. and {Ferraro}, Simone and {Foster}, Allen and {Freundt}, Rodrigo and {Fuzia}, Brittany and {Gallardo}, Patricio A. and {Garrido}, Xavier and {Gerbino}, Martina and {Giardiello}, Serena and {Gill}, Ajay and {Givans}, Jahmour and {Gluscevic}, Vera and {Goldstein}, Samuel and {Golec}, Joseph E. and {Gong}, Yulin and {Guan}, Yilun and {Halpern}, Mark and {Harrison}, Ian and {Hasselfield}, Matthew and {Healy}, Erin and {Henderson}, Shawn and {Hensley}, Brandon and {Herv{\'\i}as-Caimapo}, Carlos and {Hill}, J. Colin and {Hilton}, Gene C. and {Hilton}, Matt and {Hincks}, Adam D. and {Hlo{\v{z}}ek}, Ren{\'e}e and {Ho}, Shuay-Pwu Patty and {Hood}, John and {Hornecker}, Erika and {Huber}, Zachary B. and {Hubmayr}, Johannes and {Huffenberger}, Kevin M. and {Hughes}, John P. and {Ikape}, Margaret and {Irwin}, Kent and {Isopi}, Giovanni and {Joshi}, Neha and {Keller}, Ben and {Kim}, Joshua and {Knowles}, Kenda and {Koopman}, Brian J. and {Kosowsky}, Arthur and {Kramer}, Darby and {Kusiak}, Aleksandra and {Lagu{\"e}}, Alex and {Lakey}, Victoria and {Lee}, Eunseong and {Li}, Yaqiong and {Li}, Zack and {Limon}, Michele and {Lokken}, Martine and {Lungu}, Marius and {MacCrann}, Niall and {MacInnis}, Amanda and {Madhavacheril}, Mathew S. and {Maldonado}, Diego and {Maldonado}, Felipe and {Mallaby-Kay}, Maya and {Marques}, Gabriela A. and {van Marrewijk}, Joshiwa and {McCarthy}, Fiona and {McMahon}, Jeff and {Mehta}, Yogesh and {Menanteau}, Felipe and {Moodley}, Kavilan and {Morris}, Thomas W. and {Mroczkowski}, Tony and {Naess}, Sigurd and {Namikawa}, Toshiya and {Nati}, Federico and {Nerval}, Simran K. and {Newburgh}, Laura and {Nicola}, Andrina and {Niemack}, Michael D. and {Nolta}, Michael R. and {Orlowski-Scherer}, John and {Pagano}, Luca and {Page}, Lyman A. and {Pandey}, Shivam and {Partridge}, Bruce and {Perez Sarmiento}, Karen and {Prince}, Heather and {Puddu}, Roberto and {Qu}, Frank J. and {Ragavan}, Damien C. and {Ried Guachalla}, Bernardita and {Rogers}, Keir K. and {Rojas}, Felipe and {Sakuma}, Tai and {Schaan}, Emmanuel and {Schmitt}, Benjamin L. and {Sehgal}, Neelima and {Shaikh}, Shabbir and {Sherwin}, Blake D. and {Sierra}, Carlos and {Sievers}, Jon and {Sif{\'o}n}, Crist{\'o}bal and {Simon}, Sara and {Sonka}, Rita and {Spergel}, David N. and {Staggs}, Suzanne T. and {Storer}, Emilie and {Surrao}, Kristen and {Switzer}, Eric R. and {Tampier}, Niklas and {Thornton}, Robert and {Trac}, Hy and {Tucker}, Carole and {Ullom}, Joel and {Vale}, Leila R. and {Van Engelen}, Alexander and {Van Lanen}, Jeff and {Vargas}, Cristian and {Vavagiakis}, Eve M. and {Wagoner}, Kasey and {Wang}, Yuhan and {Wenzl}, Lukas and {Wollack}, Edward J. and {Zheng}, Kaiwen and {The Atacama Cosmology Telescope collaboration}},
        title = "{The Atacama Cosmology Telescope: DR6 power spectra, likelihoods and {\ensuremath{\Lambda}}CDM parameters}",
      journal = {JCAP},
         year = 2025,
        month = nov,
       volume = {2025},
       number = {11},
          eid = {062},
        pages = {062},
          doi = {10.1088/1475-7516/2025/11/062},
archivePrefix = {arXiv},
       eprint = {2503.14452},
 primaryClass = {astro-ph.CO},
       adsurl = {https://ui.adsabs.harvard.edu/abs/2025JCAP...11..062L}
}

@ARTICLE{Hervias-Caimapo:2025,
       author = {{Herv{\'\i}as-Caimapo}, Carlos and {Cukierman}, Ari J. and {Diego-Palazuelos}, Patricia and {Huffenberger}, Kevin M. and {Clark}, Susan E.},
        title = "{Modeling parity-violating spectra in Galactic dust polarization with filaments and its applications to cosmic birefringence searches}",
      journal = {PhRvD},
         year = 2025,
        month = apr,
       volume = {111},
       number = {8},
          eid = {083532},
        pages = {083532},
          doi = {10.1103/PhysRevD.111.083532},
archivePrefix = {arXiv},
       eprint = {2408.06214},
 primaryClass = {astro-ph.CO},
       adsurl = {https://ui.adsabs.harvard.edu/abs/2025PhRvD.111h3532H}
}

@ARTICLE{Ritacco:2024,
       author = {{Ritacco}, A. and {Bizzarri}, L. and {Savorgnano}, S. and {Boulanger}, F. and {P{\'e}rault}, M. and {Treuttel}, J. and {Morfin}, P. and {Catalano}, A. and {Darson}, D. and {Ponthieu}, N. and {Feret}, A. and {Maffei}, B. and {Chahadih}, A. and {Pisano}, G. and {Zannoni}, M. and {Nati}, F. and {Mac{\'\i}as-P{\'e}rez}, J.~F. and {Cuttaia}, F. and {Terenzi}, L. and {Monfardini}, A. and {Calvo}, M. and {Murgia}, M. and {Ortu}, P. and {Pisanu}, T. and {Aumont}, J. and {Errard}, J. and {Leclercq}, S. and {Migliaccio}, M.},
        title = "{Absolute Reference for Microwave Polarization Experiments. The COSMOCal Project and its Proof of Concept}",
      journal = {PASP},
         year = 2024,
        month = nov,
       volume = {136},
       number = {11},
          eid = {115001},
        pages = {115001},
          doi = {10.1088/1538-3873/ad8aed},
archivePrefix = {arXiv},
       eprint = {2405.12135},
 primaryClass = {astro-ph.CO},
       adsurl = {https://ui.adsabs.harvard.edu/abs/2024PASP..136k5001R}
}

@ARTICLE{Ghigna:2024,
       author = {{Ghigna}, T. and {Adler}, A. and {Aizawa}, K. and {Akamatsu}, H. and {Akizawa}, R. and {Allys}, E. and {Anand}, A. and {Aumont}, J. and {Austermann}, J. and {Azzoni}, S. and {Baccigalupi}, C. and {Ballardini}, M. and {Banday}, A.~J. and {Barreiro}, R.~B. and {Bartolo}, N. and {Basak}, S. and {Basyrov}, A. and {Beckman}, S. and {Bersanelli}, M. and {Bortolami}, M. and {Bouchet}, F. and {Brinckmann}, T. and {Campeti}, P. and {Carinos}, E. and {Carones}, A. and {Casas}, F.~J. and {Cheung}, K. and {Chinone}, Y. and {Clermont}, L. and {Columbro}, F. and {Coppolecchia}, A. and {Curtis}, D. and {de Bernardis}, P. and {de Haan}, T. and {de la Hoz}, E. and {De Petris}, M. and {Della Torre}, S. and {Delle Monache}, G. and {Di Giorgi}, E. and {Dickinson}, C. and {Diego-Palazuelos}, P. and {D{\'\i}az Garc{\'\i}a}, J.~J. and {Dobbs}, M. and {Dotani}, T. and {D'Alessandro}, G. and {Eriksen}, H.~K. and {Errard}, J. and {Essinger-Hileman}, T. and {Farias}, N. and {Ferreira}, E. and {Franceschet}, C. and {Fuskeland}, U. and {Galloni}, G. and {Galloway}, M. and {Ganga}, K. and {Gerbino}, M. and {Gervasi}, M. and {G{\'e}nova-Santos}, R.~T. and {Giardiello}, S. and {Gimeno-Amo}, C. and {Gjerl{\o}w}, E. and {Gonz{\'a}lez Gonz{\'a}lez}, R. and {Grandsire}, L. and {Gruppuso}, A. and {Halverson}, N.~W. and {Hargrave}, P. and {Harper}, S.~E. and {Hazumi}, M. and {Henrot-Versill{\'e}}, S. and {Hergt}, L.~T. and {Herranz}, D. and {Hivon}, E. and {Hlozek}, R.~A. and {Hoang}, T.~D. and {Hubmayr}, J. and {Ichiki}, K. and {Ikuma}, K. and {Ishino}, H. and {Jaehnig}, G. and {Jost}, B. and {Kohri}, K. and {Konishi}, K. and {Lamagna}, L. and {Lattanzi}, M. and {Leloup}, C. and {Levrier}, F. and {Lonappan}, A.~I. and {Luzzi}, G. and {Macias-Perez}, J. and {Maffei}, B. and {Marchitelli}, E. and {Mart{\'\i}nez-Gonz{\'a}lez}, E. and {Masi}, S. and {Matarrese}, S. and {Matsumura}, T. and {Micheli}, S. and {Migliaccio}, M. and {Monelli}, M. and {Montier}, L. and {Morgante}, G. and {Mousset}, L. and {Nagano}, Y. and {Nagata}, R. and {Natoli}, P. and {Novelli}, A. and {Noviello}, F. and {Obata}, I. and {Occhiuzzi}, A. and {Odagiri}, K. and {Omae}, R. and {Pagano}, L. and {Paiella}, A. and {Paoletti}, D. and {Pascual-Cisneros}, G. and {Patanchon}, G. and {Pavlidou}, V. and {Piacentini}, F. and {Piat}, M. and {Piccirilli}, G. and {Pinchera}, M. and {Pisano}, G. and {Porcelli}, L. and {Raffuzzi}, N. and {Raum}, C. and {Remazeilles}, M. and {Ritacco}, A. and {Rubino-Martin}, J. and {Ruiz-Granda}, M. and {Sakurai}, Y. and {Savini}, G. and {Scott}, D. and {Sekimoto}, Y. and {Shiraishi}, M. and {Signorelli}, G. and {Stever}, S.~L. and {Sullivan}, R.~M. and {Suzuki}, A. and {Takaku}, R. and {Takakura}, H. and {Takakura}, S. and {Tartari}, Y. Takase. A. and {Tassis}, K. and {Thompson}, K.~L. and {Tomasi}, M. and {Tristram}, M. and {Tucker}, C. and {Vacher}, L. and {van Tent}, B. and {Vielva}, P. and {Watanuki}, K. and {Wehus}, I.~K. and {Westbrook}, B. and {Weymann-Despres}, G. and {Winter}, B. and {Wollack}, E.~J. and {Zacchei}, A. and {Zannoni}, M. and {Zhou}, Y. and {the LiteBIRD Collaboration}},
        title = "{The LiteBIRD mission to explore cosmic inflation}",
      journal = {arXiv e-prints},
         year = 2024,
        month = jun,
          eid = {arXiv:2406.02724},
        pages = {arXiv:2406.02724},
          doi = {10.48550/arXiv.2406.02724},
archivePrefix = {arXiv},
       eprint = {2406.02724},
 primaryClass = {astro-ph.IM},
       adsurl = {https://ui.adsabs.harvard.edu/abs/2024arXiv240602724G}
}

@ARTICLE{Abitbol:2025,
       author = {{Abitbol}, M. and {Abril-Cabezas}, I. and {Adachi}, S. and {Ade}, P. and {Adler}, A.~E. and {Agrawal}, P. and {Aguirre}, J. and {Ahmed}, Z. and {Aiola}, S. and {Alford}, T. and {Ali}, A. and {Alonso}, D. and {Alvarez}, M.~A. and {An}, R. and {Arnold}, K. and {Ashton}, P. and {Atkins}, Z. and {Austermann}, J. and {Azzoni}, S. and {Baccigalupi}, C. and {Baleato Lizancos}, A. and {Barron}, D. and {Barry}, P. and {Bartlett}, J. and {Battaglia}, N. and {Battye}, R. and {Baxter}, E. and {Bazarko}, A. and {Beall}, J.~A. and {Bean}, R. and {Beck}, D. and {Beckman}, S. and {Begin}, J. and {Beheshti}, A. and {Beringue}, B. and {Bhandarkar}, T. and {Bhimani}, S. and {Bianchini}, F. and {Biermann}, E. and {Biquard}, S. and {Bixler}, B. and {Boada}, S. and {Boettger}, D. and {Bolliet}, B. and {Bond}, J.~R. and {Borrill}, J. and {Borrow}, J. and {Braithwaite}, C. and {Brien}, T.~L.~R. and {Brown}, M.~L. and {Bruno}, S.~M. and {Bryan}, S. and {Bustos}, R. and {Cai}, H. and {Calabrese}, E. and {Calafut}, V. and {Carl}, F.~M. and {Carones}, A. and {Carron}, J. and {Challinor}, A. and {Chanial}, P. and {Chen}, N. and {Cheung}, K. and {Chiang}, B. and {Chinone}, Y. and {Chluba}, J. and {Cho}, H.~S. and {Choi}, S.~K. and {Chu}, M. and {Clancy}, J. and {Clark}, S.~E. and {Clarke}, P. and {Cleary}, J. and {Clements}, D.~L. and {Connors}, J. and {Contaldi}, C. and {Coppi}, G. and {Corbett}, L. and {Cothard}, N.~F. and {Coulton}, W. and {Crowley}, K.~D. and {Crowley}, K.~T. and {Cukierman}, A. and {D'Ewart}, J.~M. and {Dachlythra}, K. and {Datta}, R. and {Day-Weiss}, S. and {de Haan}, T. and {Devlin}, M. and {Di Mascolo}, L. and {Dicker}, S. and {Dober}, B. and {Doux}, C. and {Dow}, P. and {Doyle}, S. and {Duell}, C.~J. and {Duff}, S.~M. and {Duivenvoorden}, A.~J. and {Dunkley}, J. and {Dutcher}, D. and {D{\"u}nner}, R. and {Edenton}, M. and {El Bouhargani}, H. and {Errard}, J. and {Fabbian}, G. and {Fanfani}, V. and {Farren}, G.~S. and {Fergusson}, J. and {Ferraro}, S. and {Flauger}, R. and {Foster}, A. and {Freese}, K. and {Frisch}, J.~C. and {Frolov}, A. and {Fuller}, G. and {Galitzki}, N. and {Gallardo}, P.~A. and {Galvez Ghersi}, J.~T. and {Ganga}, K. and {Gao}, J. and {Garrido}, X. and {Gawiser}, E. and {Gerbino}, M. and {Gerras}, R. and {Giardiello}, S. and {Gill}, A. and {Gilles}, V. and {Giri}, U. and {Gleave}, E. and {Gluscevic}, V. and {Goeckner-Wald}, N. and {Golec}, J.~E. and {Gordon}, S. and {Gralla}, M. and {Gratton}, S. and {Green}, D. and {Groh}, J.~C. and {Groppi}, C. and {Guan}, Y. and {Gupta}, N. and {Gudmundsson}, J.~E. and {Hagstotz}, S. and {Hargrave}, P. and {Haridas}, S. and {Harrington}, K. and {Harrison}, I. and {Hasegawa}, M. and {Hasselfield}, M. and {Haynes}, V. and {Hazumi}, M. and {He}, A. and {Healy}, E. and {Henderson}, S.~W. and {Hensley}, B.~S. and {Hertig}, E. and {Herv{\'\i}as-Caimapo}, C. and {Higuchi}, M. and {Hill}, C.~A. and {Hill}, J.~C. and {Hilton}, G. and {Hilton}, M. and {Hincks}, A.~D. and {Hinshaw}, G. and {Hlo{\v{z}}ek}, R. and {Ho}, A.~Y.~Q. and {Ho}, S. and {Ho}, S.~P. and {Hoang}, T.~D. and {Hoh}, J. and {Hornecker}, E. and {Hornsby}, A.~L. and {Hotinli}, S.~C. and {Huang}, Z. and {Huber}, Z.~B. and {Hubmayr}, J. and {Huffenberger}, K. and {Hughes}, J.~P. and {Idicherian Lonappan}, A. and {Ikape}, M. and {Irwin}, K. and {Iuliano}, J. and {Jaffe}, A.~H. and {Jain}, B. and {Jense}, H.~T. and {Jeong}, O. and {Johnson}, A. and {Johnson}, B.~R. and {Johnson}, M. and {Jones}, M. and {Jost}, B. and {Kaneko}, D. and {Karpel}, E.~D. and {Kasai}, Y. and {Katayama}, N. and {Keating}, B. and {Keller}, B. and {Keskitalo}, R. and {Kim}, J. and {Kisner}, T. and {Kiuchi}, K.},
        title = "{The Simons Observatory: science goals and forecasts for the enhanced Large Aperture Telescope}",
      journal = {JCAP},
         year = 2025,
        month = aug,
       volume = {2025},
       number = {8},
          eid = {034},
        pages = {034},
          doi = {10.1088/1475-7516/2025/08/034},
archivePrefix = {arXiv},
       eprint = {2503.00636},
 primaryClass = {astro-ph.IM},
       adsurl = {https://ui.adsabs.harvard.edu/abs/2025JCAP...08..034A}
}

@ARTICLE{Hervias-Caimapo:2022,
       author = {{Herv{\'\i}as-Caimapo}, Carlos and {Huffenberger}, Kevin M.},
        title = "{Full-sky, Arcminute-scale, 3D Models of Galactic Microwave Foreground Dust Emission Based on Filaments}",
      journal = {ApJ},
         year = 2022,
        month = mar,
       volume = {928},
       number = {1},
          eid = {65},
        pages = {65},
          doi = {10.3847/1538-4357/ac54b2},
archivePrefix = {arXiv},
       eprint = {2107.08317},
 primaryClass = {astro-ph.CO},
       adsurl = {https://ui.adsabs.harvard.edu/abs/2022ApJ...928...65H}
}

@ARTICLE{Lewis:2000,
       author = {{Lewis}, Antony and {Challinor}, Anthony and {Lasenby}, Anthony},
        title = "{Efficient Computation of Cosmic Microwave Background Anisotropies in Closed Friedmann-Robertson-Walker Models}",
      journal = {ApJ},
         year = 2000,
        month = aug,
       volume = {538},
       number = {2},
        pages = {473-476},
          doi = {10.1086/309179},
archivePrefix = {arXiv},
       eprint = {astro-ph/9911177},
 primaryClass = {astro-ph},
       adsurl = {https://ui.adsabs.harvard.edu/abs/2000ApJ...538..473L}
}

@ARTICLE{Zaldarriaga:1997,
       author = {{Zaldarriaga}, Matias and {Seljak}, Uro{\v{s}}},
        title = "{All-sky analysis of polarization in the microwave background}",
      journal = {PhRvD},
         year = 1997,
        month = feb,
       volume = {55},
       number = {4},
        pages = {1830-1840},
          doi = {10.1103/PhysRevD.55.1830},
archivePrefix = {arXiv},
       eprint = {astro-ph/9609170},
 primaryClass = {astro-ph},
       adsurl = {https://ui.adsabs.harvard.edu/abs/1997PhRvD..55.1830Z}
}

@ARTICLE{Vacher:2023a,
       author = {{Vacher}, L. and {Chluba}, J. and {Aumont}, J. and {Rotti}, A. and {Montier}, L.},
        title = "{High precision modeling of polarized signals: Moment expansion method generalized to spin-2 fields}",
      journal = {A\&A},
         year = 2023,
        month = jan,
       volume = {669},
          eid = {A5},
        pages = {A5},
          doi = {10.1051/0004-6361/202243913},
archivePrefix = {arXiv},
       eprint = {2205.01049},
 primaryClass = {astro-ph.CO},
       adsurl = {https://ui.adsabs.harvard.edu/abs/2023A&A...669A...5V}
}

@ARTICLE{Vacher:2023b,
       author = {{Vacher}, L. and {Aumont}, J. and {Boulanger}, F. and {Montier}, L. and {Guillet}, V. and {Ritacco}, A. and {Chluba}, J.},
        title = "{Frequency dependence of the thermal dust E/B ratio and EB correlation: Insights from the spin-moment expansion}",
      journal = {A\&A},
         year = 2023,
        month = apr,
       volume = {672},
          eid = {A146},
        pages = {A146},
          doi = {10.1051/0004-6361/202245292},
archivePrefix = {arXiv},
       eprint = {2210.14768},
 primaryClass = {astro-ph.CO},
       adsurl = {https://ui.adsabs.harvard.edu/abs/2023A&A...672A.146V}
}

@ARTICLE{Ritacco:2023,
       author = {{Ritacco}, Alessia and {Boulanger}, Fran{\c{c}}ois and {Guillet}, Vincent and {Delouis}, Jean-Marc and {Puget}, Jean-Loup and {Aumont}, Jonathan and {Vacher}, L{\'e}o},
        title = "{Dust polarization spectral dependence from Planck HFI data. Turning point for cosmic microwave background polarization-foreground modeling}",
      journal = {A\&A},
         year = 2023,
        month = feb,
       volume = {670},
          eid = {A163},
        pages = {A163},
          doi = {10.1051/0004-6361/202244269},
archivePrefix = {arXiv},
       eprint = {2206.07671},
 primaryClass = {astro-ph.CO},
       adsurl = {https://ui.adsabs.harvard.edu/abs/2023A&A...670A.163R}
}

@ARTICLE{Diego-Palazuelos:2023,
       author = {{Diego-Palazuelos}, P. and {Mart{\'\i}nez-Gonz{\'a}lez}, E. and {Vielva}, P. and {Barreiro}, R.~B. and {Tristram}, M. and {de la Hoz}, E. and {Eskilt}, J.~R. and {Minami}, Y. and {Sullivan}, R.~M. and {Banday}, A.~J. and {G{\'o}rski}, K.~M. and {Keskitalo}, R. and {Komatsu}, E. and {Scott}, D.},
        title = "{Robustness of cosmic birefringence measurement against Galactic foreground emission and instrumental systematics}",
      journal = {JCAP},
         year = 2023,
        month = jan,
       volume = {2023},
       number = {1},
          eid = {044},
        pages = {044},
          doi = {10.1088/1475-7516/2023/01/044},
archivePrefix = {arXiv},
       eprint = {2210.07655},
 primaryClass = {astro-ph.CO},
       adsurl = {https://ui.adsabs.harvard.edu/abs/2023JCAP...01..044D}
}

@ARTICLE{Tristram:2022,
       author = {{Tristram}, M. and {Banday}, A.~J. and {G{\'o}rski}, K.~M. and {Keskitalo}, R. and {Lawrence}, C.~R. and {Andersen}, K.~J. and {Barreiro}, R.~B. and {Borrill}, J. and {Colombo}, L.~P.~L. and {Eriksen}, H.~K. and {Fernandez-Cobos}, R. and {Kisner}, T.~S. and {Mart{\'\i}nez-Gonz{\'a}lez}, E. and {Partridge}, B. and {Scott}, D. and {Svalheim}, T.~L. and {Wehus}, I.~K.},
        title = "{Improved limits on the tensor-to-scalar ratio using BICEP and Planck data}",
      journal = {PhRvD},
         year = 2022,
        month = apr,
       volume = {105},
       number = {8},
          eid = {083524},
        pages = {083524},
          doi = {10.1103/PhysRevD.105.083524},
archivePrefix = {arXiv},
       eprint = {2112.07961},
 primaryClass = {astro-ph.CO},
       adsurl = {https://ui.adsabs.harvard.edu/abs/2022PhRvD.105h3524T}
}

@ARTICLE{Minami:2020b,
       author = {{Minami}, Yuto and {Komatsu}, Eiichiro},
        title = "{Simultaneous determination of the cosmic birefringence and miscalibrated polarization angles II: Including cross-frequency spectra}",
      journal = {PTEP},
         year = 2020,
        month = oct,
       volume = {2020},
       number = {10},
          eid = {103E02},
        pages = {103E02},
          doi = {10.1093/ptep/ptaa130},
archivePrefix = {arXiv},
       eprint = {2006.15982},
 primaryClass = {astro-ph.CO},
       adsurl = {https://ui.adsabs.harvard.edu/abs/2020PTEP.2020j3E02M}
}

@ARTICLE{Minami:2020,
       author = {{Minami}, Yuto and {Komatsu}, Eiichiro},
        title = "{New Extraction of the Cosmic Birefringence from the Planck 2018 Polarization Data}",
      journal = {PhRvL},
         year = 2020,
        month = nov,
       volume = {125},
       number = {22},
          eid = {221301},
        pages = {221301},
          doi = {10.1103/PhysRevLett.125.221301},
archivePrefix = {arXiv},
       eprint = {2011.11254},
 primaryClass = {astro-ph.CO},
       adsurl = {https://ui.adsabs.harvard.edu/abs/2020PhRvL.125v1301M}
}

@ARTICLE{Krachmalnicoff:2022,
       author = {{Krachmalnicoff}, N. and {Matsumura}, T. and {de la Hoz}, E. and {Basak}, S. and {Gruppuso}, A. and {Minami}, Y. and {Baccigalupi}, C. and {Komatsu}, E. and {Mart{\'\i}nez-Gonz{\'a}lez}, E. and {Vielva}, P. and {Aumont}, J. and {Aurlien}, R. and {Azzoni}, S. and {Banday}, A.~J. and {Barreiro}, R.~B. and {Bartolo}, N. and {Bersanelli}, M. and {Calabrese}, E. and {Carones}, A. and {Casas}, F.~J. and {Cheung}, K. and {Chinone}, Y. and {Columbro}, F. and {de Bernardis}, P. and {Diego-Palazuelos}, P. and {Errard}, J. and {Finelli}, F. and {Fuskeland}, U. and {Galloway}, M. and {Genova-Santos}, R.~T. and {Gerbino}, M. and {Ghigna}, T. and {Giardiello}, S. and {Gjerl{\o}w}, E. and {Hazumi}, M. and {Henrot-Versill{\'e}}, S. and {Kisner}, T. and {Lamagna}, L. and {Lattanzi}, M. and {Levrier}, F. and {Luzzi}, G. and {Maino}, D. and {Masi}, S. and {Migliaccio}, M. and {Montier}, L. and {Morgante}, G. and {Mot}, B. and {Nagata}, R. and {Nati}, F. and {Natoli}, P. and {Pagano}, L. and {Paiella}, A. and {Paoletti}, D. and {Patanchon}, G. and {Piacentini}, F. and {Polenta}, G. and {Poletti}, D. and {Puglisi}, G. and {Remazeilles}, M. and {Rubino-Martin}, J. and {Sasaki}, M. and {Shiraishi}, M. and {Signorelli}, G. and {Stever}, S. and {Tartari}, A. and {Tristram}, M. and {Tsuji}, M. and {Vacher}, L. and {Wehus}, I.~K. and {Zannoni}, M. and {LiteBIRD Collaboration}},
        title = "{In-flight polarization angle calibration for LiteBIRD: blind challenge and cosmological implications}",
      journal = {JCAP},
         year = 2022,
        month = jan,
       volume = {2022},
       number = {1},
          eid = {039},
        pages = {039},
          doi = {10.1088/1475-7516/2022/01/039},
archivePrefix = {arXiv},
       eprint = {2111.09140},
 primaryClass = {astro-ph.CO},
       adsurl = {https://ui.adsabs.harvard.edu/abs/2022JCAP...01..039K}
}

@ARTICLE{LiteBIRDCollaboration:2023,
       author = {{LiteBIRD Collaboration} and {Allys}, E. and {Arnold}, K. and {Aumont}, J. and {Aurlien}, R. and {Azzoni}, S. and {Baccigalupi}, C. and {Banday}, A.~J. and {Banerji}, R. and {Barreiro}, R.~B. and {Bartolo}, N. and {Bautista}, L. and {Beck}, D. and {Beckman}, S. and {Bersanelli}, M. and {Boulanger}, F. and {Brilenkov}, M. and {Bucher}, M. and {Calabrese}, E. and {Campeti}, P. and {Carones}, A. and {Casas}, F.~J. and {Catalano}, A. and {Chan}, V. and {Cheung}, K. and {Chinone}, Y. and {Clark}, S.~E. and {Columbro}, F. and {D'Alessandro}, G. and {de Bernardis}, P. and {de Haan}, T. and {de la Hoz}, E. and {De Petris}, M. and {Torre}, S. Della and {Diego-Palazuelos}, P. and {Dobbs}, M. and {Dotani}, T. and {Duval}, J.~M. and {Elleflot}, T. and {Eriksen}, H.~K. and {Errard}, J. and {Essinger-Hileman}, T. and {Finelli}, F. and {Flauger}, R. and {Franceschet}, C. and {Fuskeland}, U. and {Galloway}, M. and {Ganga}, K. and {Gerbino}, M. and {Gervasi}, M. and {G{\'e}nova-Santos}, R.~T. and {Ghigna}, T. and {Giardiello}, S. and {Gjerl{\o}w}, E. and {Grain}, J. and {Grupp}, F. and {Gruppuso}, A. and {Gudmundsson}, J.~E. and {Halverson}, N.~W. and {Hargrave}, P. and {Hasebe}, T. and {Hasegawa}, M. and {Hazumi}, M. and {Henrot-Versill{\'e}}, S. and {Hensley}, B. and {Hergt}, L.~T. and {Herman}, D. and {Hivon}, E. and {Hlozek}, R.~A. and {Hornsby}, A.~L. and {Hoshino}, Y. and {Hubmayr}, J. and {Ichiki}, K. and {Iida}, T. and {Imada}, H. and {Ishino}, H. and {Jaehnig}, G. and {Katayama}, N. and {Kato}, A. and {Keskitalo}, R. and {Kisner}, T. and {Kobayashi}, Y. and {Kogut}, A. and {Kohri}, K. and {Komatsu}, E. and {Komatsu}, K. and {Konishi}, K. and {Krachmalnicoff}, N. and {Kuo}, C.~L. and {Lamagna}, L. and {Lattanzi}, M. and {Lee}, A.~T. and {Leloup}, C. and {Levrier}, F. and {Linder}, E. and {Luzzi}, G. and {Macias-Perez}, J. and {Maciaszek}, T. and {Maffei}, B. and {Maino}, D. and {Mandelli}, S. and {Mart{\'\i}nez-Gonz{\'a}lez}, E. and {Masi}, S. and {Massa}, M. and {Matarrese}, S. and {Matsuda}, F.~T. and {Matsumura}, T. and {Mele}, L. and {Migliaccio}, M. and {Minami}, Y. and {Moggi}, A. and {Montgomery}, J. and {Montier}, L. and {Morgante}, G. and {Mot}, B. and {Nagano}, Y. and {Nagasaki}, T. and {Nagata}, R. and {Nakano}, R. and {Namikawa}, T. and {Nati}, F. and {Natoli}, P. and {Nerval}, S. and {Noviello}, F. and {Odagiri}, K. and {Oguri}, S. and {Ohsaki}, H. and {Pagano}, L. and {Paiella}, A. and {Paoletti}, D. and {Passerini}, A. and {Patanchon}, G. and {Piacentini}, F. and {Piat}, M. and {Pisano}, G. and {Polenta}, G. and {Poletti}, D. and {Prouv{\'e}}, T. and {Puglisi}, G. and {Rambaud}, D. and {Raum}, C. and {Realini}, S. and {Reinecke}, M. and {Remazeilles}, M. and {Ritacco}, A. and {Roudil}, G. and {Rubino-Martin}, J.~A. and {Russell}, M. and {Sakurai}, H. and {Sakurai}, Y. and {Sasaki}, M. and {Scott}, D. and {Sekimoto}, Y. and {Shinozaki}, K. and {Shiraishi}, M. and {Shirron}, P. and {Signorelli}, G. and {Spinella}, F. and {Stever}, S. and {Stompor}, R. and {Sugiyama}, S. and {Sullivan}, R.~M. and {Suzuki}, A. and {Svalheim}, T.~L. and {Switzer}, E. and {Takaku}, R. and {Takakura}, H. and {Takase}, Y. and {Tartari}, A. and {Terao}, Y. and {Thermeau}, J. and {Thommesen}, H. and {Thompson}, K.~L. and {Tomasi}, M. and {Tominaga}, M. and {Tristram}, M. and {Tsuji}, M. and {Tsujimoto}, M. and {Vacher}, L. and {Vielva}, P. and {Vittorio}, N. and {Wang}, W. and {Watanuki}, K. and {Wehus}, I.~K. and {Weller}, J. and {Westbrook}, B. and {Wilms}, J. and {Winter}, B. and {Wollack}, E.~J. and {Yumoto}, J. and {Zannoni}, M. and {Collaboration LiteB I R D}},
        title = "{Probing cosmic inflation with the LiteBIRD cosmic microwave background polarization survey}",
      journal = {PTEP},
         year = 2023,
        month = apr,
       volume = {2023},
       number = {4},
          eid = {042F01},
        pages = {042F01},
          doi = {10.1093/ptep/ptac150},
archivePrefix = {arXiv},
       eprint = {2202.02773},
 primaryClass = {astro-ph.IM},
       adsurl = {https://ui.adsabs.harvard.edu/abs/2023PTEP.2023d2F01L}
}

@ARTICLE{Martire:2022,
       author = {{Martire}, F.~A. and {Barreiro}, R.~B. and {Mart{\'\i}nez-Gonz{\'a}lez}, E.},
        title = "{Characterization of the polarized synchrotron emission from Planck and WMAP data}",
      journal = {JCAP},
         year = 2022,
        month = apr,
       volume = {2022},
       number = {4},
          eid = {003},
        pages = {003},
          doi = {10.1088/1475-7516/2022/04/003},
archivePrefix = {arXiv},
       eprint = {2110.12803},
 primaryClass = {astro-ph.CO},
       adsurl = {https://ui.adsabs.harvard.edu/abs/2022JCAP...04..003M}
}

@ARTICLE{Kamionkowski:2016,
       author = {{Kamionkowski}, Marc and {Kovetz}, Ely D.},
        title = "{The Quest for B Modes from Inflationary Gravitational Waves}",
      journal = {ARA\&A},
         year = 2016,
        month = sep,
       volume = {54},
        pages = {227-269},
          doi = {10.1146/annurev-astro-081915-023433},
archivePrefix = {arXiv},
       eprint = {1510.06042},
 primaryClass = {astro-ph.CO},
       adsurl = {https://ui.adsabs.harvard.edu/abs/2016ARA&A..54..227K}
}
\bibliographystyle{spiebib} 

\end{document}